\documentclass[ALICE,manyauthors]{cernphprep}
\usepackage[comma,square,numbers,sort&compress]{natbib}
\usepackage{hyperref}
\usepackage{lineno}
\usepackage{multirow}
\usepackage{makecell}
\usepackage{tabularx}
\usepackage{xcolor}
\usepackage{xspace}
\usepackage{float}
\usepackage[T1]{fontenc}
\usepackage{orcidlink}

\begin{document}
%

\newcommand{\pp}           {pp\xspace}
\newcommand{\ppbar}        {\mbox{$\mathrm {p\overline{p}}$}\xspace}
\newcommand{\XeXe}         {\mbox{Xe--Xe}\xspace}
\newcommand{\PbPb}         {\mbox{Pb--Pb}\xspace}
\newcommand{\pA}           {\mbox{pA}\xspace}
\newcommand{\pPb}          {\mbox{p--Pb}\xspace}
\newcommand{\AuAu}         {\mbox{Au--Au}\xspace}
\newcommand{\dAu}          {\mbox{d--Au}\xspace}

\newcommand{\s}            {\ensuremath{\sqrt{s}}\xspace}
\newcommand{\snn}          {\ensuremath{\sqrt{s_{\mathrm{NN}}}}\xspace}
\newcommand{\pt}           {\ensuremath{p_{\rm T}}\xspace}
\newcommand{\ptrange}[2]   {\mbox{$#1 < {p_{\rm T}} < #2$}}
\newcommand{\meanpt}       {$\langle p_{\mathrm{T}}\rangle$\xspace}
\newcommand{\ycms}         {\ensuremath{y_{\rm CMS}}\xspace}
\newcommand{\ylab}         {\ensuremath{y_{\rm lab}}\xspace}
\newcommand{\etarange}[2]  {\mbox{$#1 < \eta < #2$}}
\newcommand{\etarangeAb}[1]{\mbox{$\left | \eta \right |~<~#1$}}
\newcommand{\yrangeAb}[1]  {\mbox{$\left | y \right |~<~#1$}}
\newcommand{\dndy}         {\ensuremath{\mathrm{d}N_\mathrm{ch}/\mathrm{d}y}\xspace}
\newcommand{\dndeta}       {\ensuremath{\mathrm{d}N_\mathrm{ch}/\mathrm{d}\eta}\xspace}
\newcommand{\avdndeta}     {\ensuremath{\langle\dndeta\rangle}\xspace}
\newcommand{\dNdy}         {\ensuremath{\mathrm{d}N_\mathrm{ch}/\mathrm{d}y}\xspace}
\newcommand{\Npart}        {\ensuremath{N_\mathrm{part}}\xspace}
\newcommand{\Ncoll}        {\ensuremath{N_\mathrm{coll}}\xspace}
\newcommand{\dEdx}         {\ensuremath{\textrm{d}E/\textrm{d}x}\xspace}
\newcommand{\RpPb}         {\ensuremath{R_{\rm pPb}}\xspace}

\newcommand{\nineH}        {$\sqrt{s}~=~0.9$~Te\kern-.1emV\xspace}
\newcommand{\seven}        {$\sqrt{s}~=~7$~Te\kern-.1emV\xspace}
\newcommand{\twoH}         {$\sqrt{s}~=~0.2$~Te\kern-.1emV\xspace}
\newcommand{\twosevensix}  {$\sqrt{s}~=~2.76$~Te\kern-.1emV\xspace}
\newcommand{\five}         {$\sqrt{s}~=~5.02$~Te\kern-.1emV\xspace}
\newcommand{\twosevensixnn}{$\sqrt{s_{\mathrm{NN}}}~=~2.76$~Te\kern-.1emV\xspace}
\newcommand{\fivenn}       {$\sqrt{s_{\mathrm{NN}}}~=~5.02$~Te\kern-.1emV\xspace}
\newcommand{\LT}           {L{\'e}vy-Tsallis\xspace}
\newcommand{\GeVc}         {Ge\kern-.1emV/$c$\xspace}
\newcommand{\MeVc}         {Me\kern-.1emV/$c$\xspace}
\newcommand{\TeV}          {Te\kern-.1emV\xspace}
\newcommand{\GeV}          {Ge\kern-.1emV\xspace}
\newcommand{\MeV}          {Me\kern-.1emV\xspace}
\newcommand{\GeVmass}      {Ge\kern-.1emV/$c^2$\xspace}
\newcommand{\MeVmass}      {Me\kern-.1emV/$c^2$\xspace}
\newcommand{\lumi}         {\ensuremath{\mathcal{L}}\xspace}

\newcommand{\ITS}          {\rm{ITS}\xspace}
\newcommand{\TOF}          {\rm{TOF}\xspace}
\newcommand{\ZDC}          {\rm{ZDC}\xspace}
\newcommand{\ZDCs}         {\rm{ZDCs}\xspace}
\newcommand{\ZNA}          {\rm{ZNA}\xspace}
\newcommand{\ZNC}          {\rm{ZNC}\xspace}
\newcommand{\SPD}          {\rm{SPD}\xspace}
\newcommand{\SDD}          {\rm{SDD}\xspace}
\newcommand{\SSD}          {\rm{SSD}\xspace}
\newcommand{\TPC}          {\rm{TPC}\xspace}
\newcommand{\TRD}          {\rm{TRD}\xspace}
\newcommand{\VZERO}        {\rm{V0}\xspace}
\newcommand{\VZEROA}       {\rm{V0A}\xspace}
\newcommand{\VZEROC}       {\rm{V0C}\xspace}
\newcommand{\Vdecay} 	   {\ensuremath{V^{0}}\xspace}

\newcommand{\ee}           {\ensuremath{e^{+}e^{-}}\xspace}
\newcommand{\pip}          {\ensuremath{\pi^{+}}\xspace}
\newcommand{\pim}          {\ensuremath{\pi^{-}}\xspace}
\newcommand{\kap}          {\ensuremath{\rm{K}^{+}}\xspace}
\newcommand{\kam}          {\ensuremath{\rm{K}^{-}}\xspace}
\newcommand{\pbar}         {\ensuremath{\rm\overline{p}}\xspace}
\newcommand{\kzero}        {\ensuremath{{\rm K}^{0}_{\rm{S}}}\xspace}
\newcommand{\lmb}          {\ensuremath{\Lambda}\xspace}
\newcommand{\almb}         {\ensuremath{\overline{\Lambda}}\xspace}
\newcommand{\Om}           {\ensuremath{\Omega^-}\xspace}
\newcommand{\Mo}           {\ensuremath{\overline{\Omega}^+}\xspace}
\newcommand{\X}            {\ensuremath{\Xi^-}\xspace}
\newcommand{\Ix}           {\ensuremath{\overline{\Xi}^+}\xspace}
\newcommand{\Xis}          {\ensuremath{\Xi^{\pm}}\xspace}
\newcommand{\Oms}          {\ensuremath{\Omega^{\pm}}\xspace}
\newcommand{\degree}       {\ensuremath{^{\rm o}}\xspace}

\newcommand{\epem}         {\mbox{$\mathrm {e^{+}e^{-}}$}\xspace}
\newcommand{\ep}           {\mbox{$\mathrm{e}^{-}\mathrm{p}$}\xspace}
\newcommand{\thirteen}     {$\sqrt{s}=13$~Te\kern-.1emV\xspace}
\newcommand{\thirteensix}  {$\sqrt{s}~=~13.6$~Te\kern-.1emV\xspace}
\newcommand{\VZEROM}       {\rm{V0M}\xspace}

\newcommand{\pizero}       {\ensuremath{\pi^{0}}\xspace}
\newcommand{\Lambdab}      {\ensuremath{\Lambda_{\rm b}^{0}}\xspace}
\newcommand{\Lambdac}      {\ensuremath{\Lambda_{\rm c}^{+}}\xspace}
\newcommand{\Xib}          {\ensuremath{\Xi_{\rm b}}\xspace}
\newcommand{\Xic}          {\ensuremath{\Xi_{\rm c}}\xspace}
\newcommand{\Xicplus}      {\ensuremath{\Xi_{\rm c}^{+}}\xspace}
\newcommand{\Xiczero}      {\ensuremath{\Xi_{\rm c}^{0}}\xspace}
\newcommand{\Xiczeroplus}  {\ensuremath{\Xi_{\rm c}^{0,+}}\xspace}
\newcommand{\Sigmac}       {\ensuremath{\Sigma_{\rm c}^{0,++}\mathrm{(2455)}}\xspace}
\newcommand{\Dzero}        {\ensuremath{\rm D^{0}}\xspace}
\newcommand{\Ds}           {\ensuremath{\rm D^{+}_{s}}\xspace}
\newcommand{\Dplus}        {\ensuremath{\rm D^{+}}\xspace}
\newcommand{\Dstar}        {\ensuremath{\rm D^{*+}}\xspace}
\newcommand{\Omegac}       {\ensuremath{\Omega_{\rm c}^{0}}\xspace}
\newcommand{\nue}          {\ensuremath{\nu_{\rm e}}\xspace}
\newcommand{\eplus}        {\ensuremath{e^{+}}\xspace}

\newcommand{\acceff}       {$(\rm{Acc} \times \epsilon)$\xspace}
\newcommand{\XiePair}      {\ensuremath{\rm e\Xi}\xspace}
\newcommand{\ntrk}         {$N_{\rm tracklet}$\xspace}

\newcommand{\slfrac}[2]{\left.#1\right/#2}
\newcommand{\XicZeroToPiXi}{$\rm \Xi_{c}^{0} \rightarrow \Xi^{-} \pi^{+}$\xspace}
\newcommand{\XicZeroToXiEleNu}{$\Xiczero \rightarrow \Xi^- {\rm e}^+ \rm \nu_{\rm e}$\xspace}
\newcommand{\XicPlusToXiPiPi}{$\rm \Xi_{c}^{+} \rightarrow \Xi^{-} \pi^{+} \pi^{+}$\xspace}
\newcommand{\pvom}{$\it{p}_{\mathrm{V0M}}$\xspace}

\definecolor{mypink}{rgb}{1 0.48 0.81}
\definecolor{mySkyblue}{rgb}{0.1 0.7 1.0}
\definecolor{myGreen}{rgb}{0.07 0.79 0.25}

\begin{titlepage}
\PHyear{2025}       
\PHnumber{175}      
\PHdate{4 August}  

\title{Multiplicity dependence of \Xicplus and \Xiczero production\\in \pp collisions at \s = 13 \TeV}
\ShortTitle{Multiplicity dependence of \Xicplus and \Xiczero in \pp at 13 \TeV} 

\Collaboration{ALICE Collaboration\thanks{See Appendix~\ref{app:collab} for the list of collaboration members}}
\ShortAuthor{ALICE Collaboration} 


\begin{abstract}

The first measurement at midrapidity (\yrangeAb{0.5}) of the production yield of the strange-charm baryons \Xicplus and \Xiczero as a function of transverse momentum (\pt) in different charged-particle multiplicity classes in proton--proton collisions at \thirteen with the ALICE experiment at the LHC is reported.
The \Xicplus baryon is reconstructed via the \XicPlusToXiPiPi decay channel in the range \ptrange{4}{12} \GeVc, while the \Xiczero baryon is reconstructed via both the \XicZeroToPiXi and \XicZeroToXiEleNu decay channels in the range \ptrange{2}{12} \GeVc.
The baryon-to-meson (\Xiczeroplus/\Dzero) and the baryon-to-baryon (\Xiczeroplus/\Lambdac) production yield ratios show no significant dependence on multiplicity.
In addition, the observed yield ratios are not described by theoretical predictions that model charm-quark fragmentation based on measurements at \epem and \ep colliders, indicating differences in the charm-baryon production mechanism in pp collisions.
A comparison with different event generators and tunings, including different modelling of the hadronisation process, is also discussed.
Moreover, the branching-fraction ratio of BR(\XicZeroToXiEleNu)/BR(\XicZeroToPiXi) is measured as 0.825 $\pm$ 0.094 (stat.) $\pm$ 0.081 (syst.). This value supersedes the previous ALICE measurement, improving the statistical precision by a factor of 1.6.

\end{abstract}


\end{titlepage}
\setcounter{page}{2} 


\clearpage
\section{Introduction}\label{sec:intro}

Heavy quarks, such as charm and beauty, are produced in hard-scattering processes occurring in the early stages of ultrarelativistic proton--proton (\pp) collisions at the LHC. Therefore, measuring the production of hadrons containing at least one heavy quark provides a crucial test of perturbative Quantum Chromodynamics (pQCD) calculations.
The production cross section of heavy-flavour hadrons, generally computed using a factorisation approach, is calculated as the convolution of three factors~\cite{Collins:1985gm}: i) the Parton Distribution Functions (PDFs), which describe the probability distribution of finding quarks or gluons inside the incoming proton with a given fraction of the proton momentum; ii) the hard-scattering cross section at the partonic level, which governs the interaction between partons leading to the production of heavy quarks; and iii) the fragmentation functions, which model the transition of the heavy quark into a heavy-flavour hadron carrying a fraction of the quark momentum.
The fragmentation functions cannot be computed using the pQCD approach. They are typically constrained by measurements in \epem or \ep collisions~\cite{Braaten_1995}, assuming universal hadronisation. Measurements of heavy-flavour hadron production in pp collisions provide the essential ingredients for testing the universality of heavy-quark fragmentation functions, and ultimately the heavy-quark hadronisation process. The production yield ratios of different hadron species are an especially sensitive tool to investigate hadronisation mechanisms since the PDFs and partonic scattering cross sections are universal across all charm hadrons.

Extensive measurements of charm-hadron production including charm mesons: \Dzero, \Dplus, \Ds, \Dstar,
$\mathrm{D_{s1}^{+}}(2536)$, $\mathrm{D_{s2}^{*+}(2573)}$~\cite{ALICE:2017thy, ALICE:2019nxm, ALICE:2021mgk, ATLAS:2015igt, CMS:jhep2021_225, LHCb:2016ikn,LHCb:2015swx, ALICE:2024hkk}, and charm baryons: \Lambdac, \Sigmac, \Xiczeroplus, \Omegac~\cite{ALICE:2023sgl, ALICE:2020wfu, ALICE:2020wla, ALICE:2022exq, CMS:2019uws, ALICE:2021rzj, ALICE:2021bli, ALICE:2021psx, ALICE:2022cop}, have been performed in pp collisions at the LHC~\cite{ALICE:2022wpn}.
The measurements of charm-meson production cross sections at midrapidity in pp collisions are well described over a wide range of transverse momentum (\pt) with respect to the beam axis, by perturbative calculations at next-to-leading order with next-to-leading-log resummation, such as the general-mass variable-flavour-number scheme (GM-VFNS~\cite{Kramer:2017gct,Helenius:2018uul,Kniehl:2020szu}) and the fixed-order plus next-to-leading-log (FONLL~\cite{Cacciari:1998it, Cacciari:2012ny}) frameworks. Both calculations use fragmentation functions tuned on measurements in \epem and \ep collisions~\cite{Braaten_1995}.
However, these models significantly underestimate charm-baryon production in pp collisions~\cite{ALICE:2017thy, ALICE:2020wla}. Also, the measurements of the \Lambdac/\Dzero baryon-to-meson production yield ratio in \pp collisions at $\sqrt{s}~=~5.02$, 7, and 13 \TeV~\cite{ALICE:2020wfu,ALICE:2017thy,ALICE:2021rzj} show a strong \pt dependence, with a significant enhancement at \ptrange{1}{8} \GeVc, compared to the results obtained in \epem and \ep collisions~\cite{Braaten_1995}. This enhancement also deviates from the predictions of event generators tuned to reproduce the results in \epem collisions, such as PYTHIA~8 with the Monash tune~\cite{Skands_2014}. 
A similar deviation between pp and \epem collisions has been reported also for the beauty-quark baryon-to-meson production ratio, suggesting that the observed difference is not limited to the charm sector~\cite{LHCb:2019fns, ALICE:2023wbx}.
These observations question the assumed universality of the charm-quark hadronisation process across different collision systems and suggest that the `in-vacuum' string fragmentation alone fails to describe heavy-flavour baryon production in hadronic collisions.
Recent ALICE measurements on charm fragmentation fractions in \pp collisions at $\sqrt{s}~=~5.02$ and 13 \TeV~\cite{ALICE:2021dhb, ALICE:2023sgl} also support this conclusion.

The \Lambdac/\Dzero production ratio in \pp collisions is well described by effective model calculations and event generators implementing different hadronisation mechanisms: i) a tune of the PYTHIA~8 event generator with colour reconnection beyond the leading-colour approximation (CR-BLC~\cite{Christiansen:2015yqa}), in which baryon production is enhanced by the introduction of new topologies for colour reconnection mechanisms as the junctions; ii) the Statistical Hadronisation Model (SHM) including higher-mass excited charm baryon states, which are predicted by the Relativistic Quark Model (RQM)~\cite{SHMRQM} but have not yet been experimentally measured. The model replaces the complexity of the hadronisation process with thermo-statistical weights governed by the masses of available hadron states at a universal hadronisation temperature; and iii) models assuming a charm quark hadronisation via coalescence with other quarks produced in the collision, such as Catania~\cite{Catania1,Catania2}, Quark Combination Mode (QCM)~\cite{QCM}, and POWLANG~\cite{POWLANG}.
However, unlike the measurements of the \Lambdac/\Dzero and \Sigmac/\Dzero ratios~\cite{ALICE:2021rzj} that are both described by such models within the measured uncertainties, the production ratios for strange-charm baryons, such as \Xiczeroplus/\Dzero and \Omegac/\Dzero, are not well described by these models~\cite{ALICE:2021bli, ALICE:2022cop}.
Studying charm-hadron production as a function of charged-particle multiplicity in pp collisions can extend the knowledge about how the charm-quark hadronisation mechanism evolves from low to high particle densities. 
Recent results of multiplicity-dependent \Lambdac/\Dzero production ratios in \pp collisions~\cite{ALICE:2021npz} exhibit an increasing charm baryon-to-meson ratio with an increasing charged-particle multiplicity in the intermediate \pt region (\ptrange{2}{12} \GeVc). However, at the same time, no significant multiplicity dependence is observed for the \pt-integrated yields within the current uncertainties.
This weak multiplicity dependence of the \pt-integrated \Lambdac/\Dzero ratio is similar to that of $\Lambda$/\kzero~\cite{ALICE:2019avo}. Interestingly, a clear multiplicity dependence is observed in the multi-strange baryon-to-meson production yield ratios \X/\kzero and \Om/\kzero.
One possible explanation in the context of the Canonical Statistical Model (CSM)~\cite{VOVCHENKO2018171} is that moving from pp to Pb--Pb collisions the system volume increases, which gradually removes the canonical suppression associated with an exact charge conservation in pp collisions, leading to a global charge conservation in Pb--Pb collisions.
Therefore, measurements of strange-charm baryon production in various multiplicity classes will offer key insights into hadron production mechanisms and their evolution with multiplicity density.

\renewcommand{\arraystretch}{1.35}
\begin{table}[t]
    \centering
    \caption{The branching ratios for the relevant decay channels}
    \begin{tabular}{l|l|l|l}
    \hline
    \hline
    Decay channel & BR (\%) & Decay channel & BR (\%) \\\hline  
    \XicPlusToXiPiPi & 2.86 $\pm$ 1.21 $\pm$ 0.38~\cite{Belle:2019bgi} &
    $\Xi^{-} \rightarrow \Lambda\pi^{-}$ & 99.9 $\pm$ 0.04~\cite{ParticleDataGroup:2022ynf} \\
    \XicZeroToPiXi & 1.43 $\pm$ 0.27~\cite{ParticleDataGroup:2022ynf} &
    $\Lambda \rightarrow \mathrm{p}\pi^{-}$ & 64.1 $\pm$ 0.5~\cite{ParticleDataGroup:2022ynf} \\
    \XicZeroToXiEleNu & 1.04 $\pm$ 0.24~\cite{ParticleDataGroup:2022ynf} & - & - \\
    \hline
    \hline
    \end{tabular}
    \label{tab:BR}
\end{table}

This article reports the \pt-differential production yields of prompt \Xicplus and \Xiczero baryons -- produced either directly from a charm-quark hadronisation or from strong decays of excited charm-hadron states -- measured in pp collisions at $\sqrt{s}=13$ TeV at midrapidity (\yrangeAb{0.5}), in intervals of charged-particle multiplicity.
The \Xicplus baryon is reconstructed via the decay channel \XicPlusToXiPiPi (and charge conjugates), in the interval 4 $<$ \pt $<$ 12 \GeVc while the \Xiczero baryon is reconstructed via the decay channels \XicZeroToXiEleNu and \XicZeroToPiXi (and charge conjugates) in the interval 2 $<$ \pt $<$ 12 \GeVc. The relevant branching ratios are summarised in Table~\ref{tab:BR}. The \pt-dependent $\Xi_{\rm c}^{0, +} / \Lambda_c^+$  and $\Xi_{\rm c}^{0, +} / {\rm D}^0$ production yield ratios are also reported.
In addition, the branching fraction ratio BR(\XicZeroToXiEleNu)/BR(\XicZeroToPiXi) is measured with improved precision compared to previous results~\cite{ALICE:2021bli}, thanks to the increased statistics.

This article is organised as follows:
Section~\ref{sec:exp} describes the experimental setup and the multiplicity determination,
Section~\ref{sec:dataAna} explains the analysis details,
Section~\ref{sec:systErr} discusses the estimation of the systematic uncertainties,
Section~\ref{sec:result} presents the results and comparisons to the model calculations,
and Section~\ref{sec:summary} concludes with a summary.

\section{Experimental apparatus and data sample}\label{sec:exp}

Detailed descriptions of the ALICE apparatus and its LHC Run 2 performance are provided in Refs.~\cite{ALICE:2008ngc,ALICE:2014sbx}. In this section, the main detectors used for the analyses discussed in this article are briefly described.
The central barrel detectors are located inside a cylindrical solenoid magnet, providing a magnetic field along the beam direction with a field strength of 0.5 T. The central barrel detectors, which cover the pseudorapidity interval \etarangeAb{0.9}, are the Inner Tracking System (\ITS), the Time Projection Chamber (\TPC), and the Time-Of-Flight (\TOF) detectors, in order of radial distance from the beam axis.
At midrapidity, the \ITS and \TPC provide track and vertex reconstruction. The \TPC and \TOF are used for particle identification (PID). At forward rapidity, the \VZERO detector assembly provides multiplicity estimation and triggering, with the V0A and V0C sub-detectors covering the pseudorapidity ranges \etarange{2.8}{5.1} and \etarange{-3.7}{-1.7}, respectively.

The data samples collected during the LHC Run 2 (2016, 2017, and 2018) from \pp collisions at a centre-of-mass energy of \thirteen were used. During the data taking, two trigger conditions based on the signal amplitude recorded in the \VZERO detector were used. The minimum-bias (MB) trigger requires coincident signals in \VZEROA and \VZEROC at the same proton bunch crossing time. To enhance the selection of high-multiplicity events, a dedicated trigger, called the High Multiplicity \VZERO trigger (HMV0) was used. The HMV0 trigger condition is satisfied if the sum of signal amplitude in \VZEROA and \VZEROC (denoted as \VZEROM) is 5 times larger than the mean value in MB-triggered samples~\cite{ALICE:2021npz}. The trigger efficiencies are reported in Table~\ref{tab:multiplicity class}, as estimated in Ref.~\cite{ALICE:2020swj}.

Further offline event selections were applied to remove the contamination from beam-gas collisions and other machine-related backgrounds. These selection criteria were based on the signals from the \VZERO detector and the two innermost layers of the \ITS, which constitute the Silicon Pixel Detector (\SPD). Only events with a primary vertex within $\pm10$ cm of the nominal interaction point along the beam axis were selected. Moreover, events with multiple reconstructed primary vertices in the same proton bunch crossing (pile-up events) were excluded. In addition, a condition that requires at least one charged particle produced within the interval \etarangeAb{1} (denoted as INEL $>$ 0) was applied to reject diffractive interactions. The events satisfying the aforementioned conditions account for about 75\% of the total inelastic cross section~\cite{ALICE:2015ikl, ALICE:2020swj}. After the selections, the integrated luminosity of the analysed data sample was about $\mathcal{L}_{\mathrm{int}}$ $\approx$ 32 $\mathrm{nb}^{-1}$ for the MB trigger and $\mathcal{L}_{\mathrm{int}}$ $\approx$ 7.7 $\mathrm{pb}^{-1}$ for the HMV0 trigger~\cite{ALICE:2021npz}.

The INEL $>$ 0 events were categorised into three classes based on the charged-particle multiplicity, which was estimated from the percentile distribution of \VZEROM (\pvom).
Note that a small (large) value of \pvom corresponds to high (low) charged-particle multiplicity.
The \pvom intervals were converted to the average charged-particle multiplicity in \etarangeAb{0.5} ($\left< \dndeta \right>_{|\eta|<0.5}$), as explained in Refs.~\cite{ALICE:2021npz, ALICE:2020swj}. In Table~\ref{tab:multiplicity class}, the event classes and corresponding \pvom intervals, along with the $\left< \dndeta \right>_{|\eta|<0.5}$ values, are summarised. The high-multiplicity (0--0.1\%) events were collected with the HMV0 trigger, while the other multiplicity classes (0--100\%, 0.1--30\%, and 30--100\%) were collected with the MB trigger.

\renewcommand{\arraystretch}{1.35}
\begin{table} [t]
    \centering
    \caption{Multiplicity classes based on \pvom percentile and the corresponding $\left< \dndeta \right>_{|\eta|<0.5}$~\cite{ALICE:2021npz, ALICE:2020swj}}
    \vspace{0.2cm}
    \begin{tabular}{c|c|c|c|c}
    \hline
    \hline
    \thead{\normalsize Description} & \thead{\normalsize Trigger} &
    \thead{\normalsize Trigger efficiency\\[2pt](\normalsize $\epsilon_{\mathrm{trigger}}$)} &
    \thead{\normalsize Multiplicity class\\[2pt](\normalsize \pvom[\%])} &
    \thead{\normalsize $\left< \dndeta \right>_{|\eta|<0.5}$} \\\hline
    INEL $>$ 0   & MB   & 0.920 $\pm$ 0.003 & [0, 100]  & 6.93 $\pm$ 0.09 \\
    High multiplicity & HMV0 & 1 (negl. unc.)    & [0, 0.1]  & 31.53 $\pm$ 0.38 \\
    Intermediate multiplicity & MB   & 0.997 $\pm$ 0.001 & [0.1, 30] & 13.81 $\pm$ 0.14 \\
    Low multiplicity & MB   & 0.897 $\pm$ 0.013 & [30, 100] & 4.41 $\pm$ 0.05  \\
    \hline
    \hline
    \end{tabular}
    \label{tab:multiplicity class}
\end{table}

\section{Data analysis}\label{sec:dataAna}

In this article, \Xicplus and \Xiczero with their charge conjugates were analysed using hadronic decay channels (\XicPlusToXiPiPi and \XicZeroToPiXi) and a semileptonic decay channel (\XicZeroToXiEleNu) at midrapidity ($|y|$ $<$ 0.5). The \X baryons appearing in these decay chains were reconstructed from $\Xi^{-}\rightarrow\Lambda\pi^{-}$, followed by $\Lambda\rightarrow\mathrm{p}\pi^{-}$~\cite{ALICE:2021bli}.
In the analyses, Monte Carlo (MC) simulations with the PYTHIA 8~\cite{Sjostrand:2014zea} event generator were used for various purposes, such as machine learning model training, acceptance-times-efficiency \acceff correction, unfolding, and template fit. The charm hadrons generated via PYTHIA 8 were forced to decay via the specific decay channel of interest in the analyses presented in this article. The interactions between the generated particles and the detectors were modelled using the GEANT3 package~\cite{Brun:1082634} and the detector geometry and conditions during the data taking were reproduced.

\begin{figure}[b]
    \centering
    \includegraphics[width=0.475\textwidth]{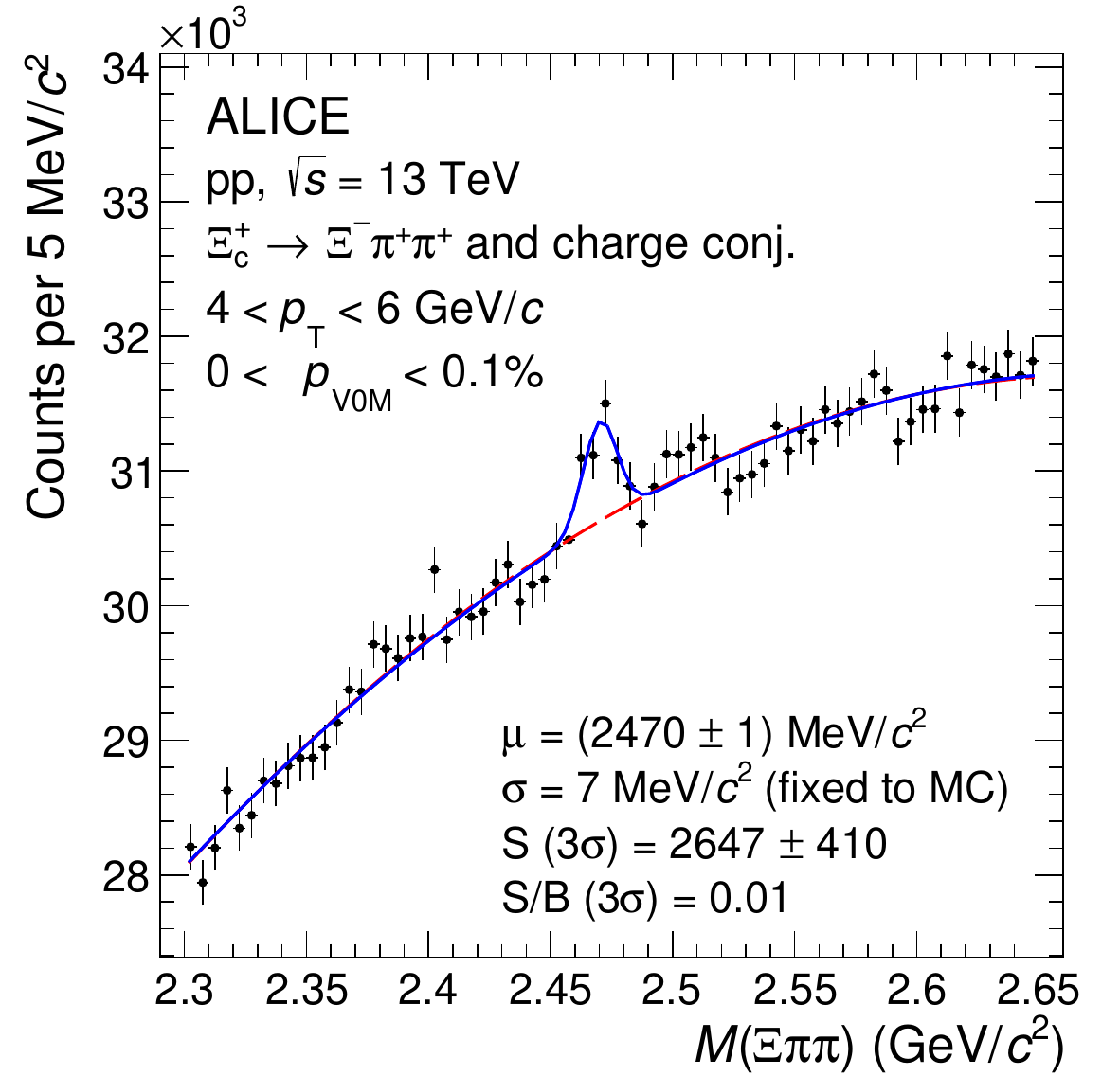}
      \includegraphics[width=0.475\textwidth]{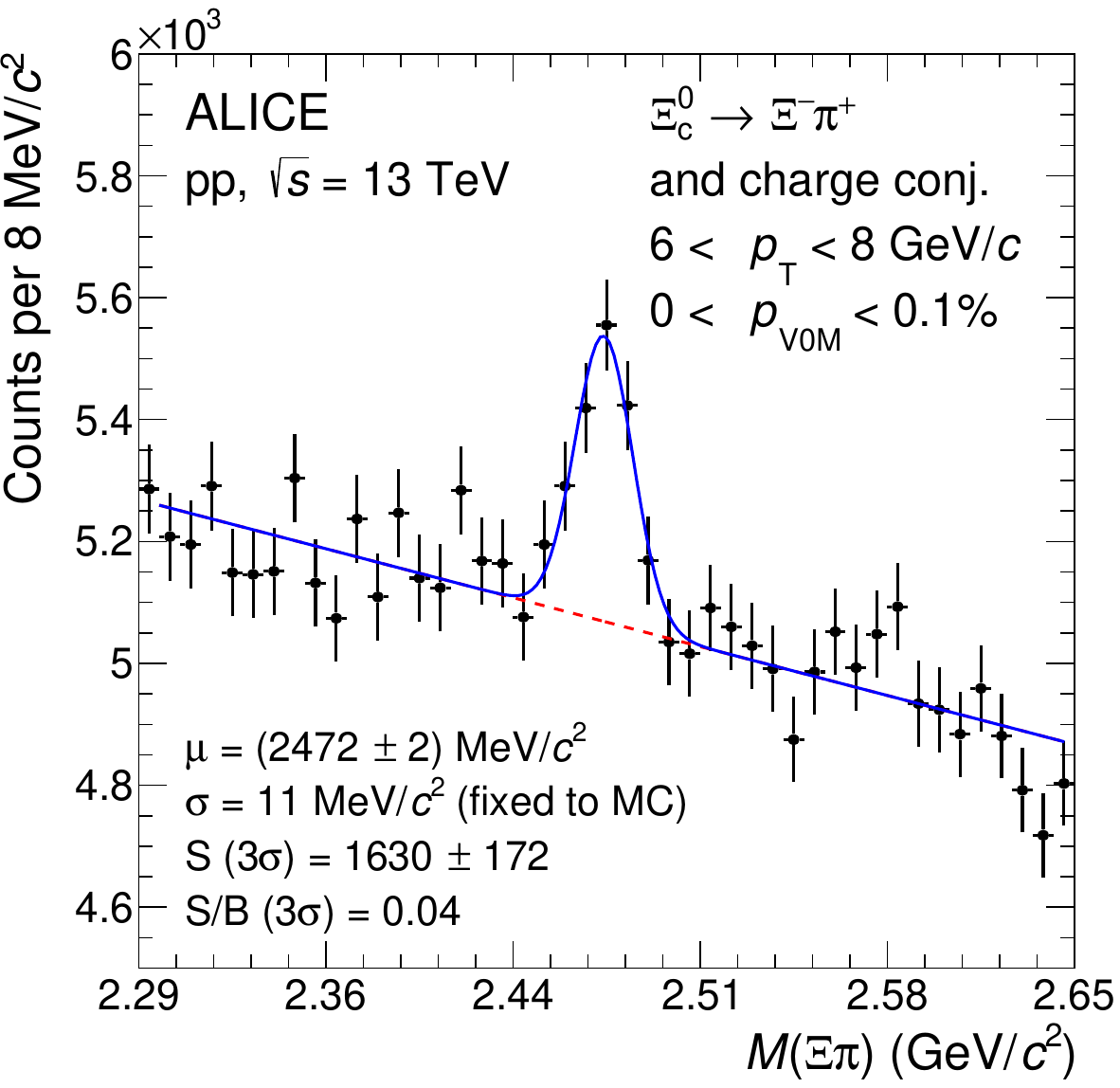}
    \caption{Invariant-mass distributions of signal candidates for the hadronic decays of \Xicplus in 4 $<$ \pt $<$ 6 \GeVc (left), and \Xiczero in 6 $<$ \pt $<$ 8 \GeVc (right), in the high-multiplicity class. The blue solid curve shows the total fit, and the red dashed curve shows the combinatorial background.}
    \label{fig:Invamass}
\end{figure}

To reconstruct the \Xiczeroplus via their hadronic decay channels, a $\Xi^{-}$ baryon candidate was paired with either one or two tracks, as described in Refs.~\cite{ALICE:2021bli,ALICE:2023sgl}. The tracks were required to have at least 70 crossed pad rows in the \TPC and at least 3 hits in the \ITS. The identification of pions and protons was achieved by requiring the specific energy loss \dEdx of the track and time-of-flight to be compatible with the expected values in units of the detector resolution ($n_{\sigma}^{\rm det}$) within three standard deviations. Additional track quality selections were applied, as described in Ref.~\cite{ALICE:2023sgl}.
Additionally, for \Xiczero candidates, the Kalman Filter Particle tracking algorithm~\cite{Kisel:2012xia} was used to reconstruct the \XicZeroToPiXi decay channel.

A Boosted Decision Tree (BDT) algorithm, implemented using the XGBoost library~\cite{Chen:2016btl} was adopted to separate the combinatorial background from the signal.
The binary BDT classifiers were independently trained for each \pt interval with the sample extracted from the inclusive MB sample~\cite{ALICE:2021mgk,ALICE:2022exq,ALICE:2021kfc,ALICE:2022cop}. The signal sample was obtained from MC simulations and the background sample was collected from data by selecting reconstructed \Xiczeroplus candidates located at least 6$\sigma$ away from the invariant-mass signal peak, where $\sigma$ corresponds to the width of the signal peak obtained from MC simulations.
The BDT models were trained using variables related to PID and the decay topology of \Xiczeroplus baryons, such as the distance of closest approach (DCA) between the $\Xi^{-}$ baryon and the pion coming from a \Xiczeroplus candidate, the DCA between the primary vertex and daughter particles, the pointing angle between the reconstructed momentum vector of \Xiczeroplus baryon and the vector that connects the primary and secondary vertices, and the reconstructed invariant mass of the $\Xi$ and $\Lambda$.
The output score from the trained BDT model allows each candidate to be classified with the probability of being a true signal. In each \pt interval, the threshold value of the BDT score was chosen to maximise the expected statistical significance of the signal, according to the procedure described in Ref.~\cite{ALICE:2020wfu}. The \Xiczeroplus signal was measured via a binned maximum-likelihood fit of the candidate invariant-mass distribution in each multiplicity class in the \pt intervals \ptrange{4}{12} \GeVc for \Xicplus and \ptrange{2}{12} \GeVc for the \Xiczero. A Gaussian function was used to describe the signal, and its width ($\sigma$) was fixed to the value observed in MC simulations to improve the stability of the fit. The background was modelled with a first- or second-order polynomial, depending on the shape of the background in each \pt and multiplicity interval. The statistical significance of the obtained \Xiczeroplus signal ranged from 3.2$\sigma$ to 8.4$\sigma$. Some examples of these fits are presented in Fig.~\ref{fig:Invamass}.

\begin{figure}[b]
    \centering
    \includegraphics[width=0.475\textwidth]{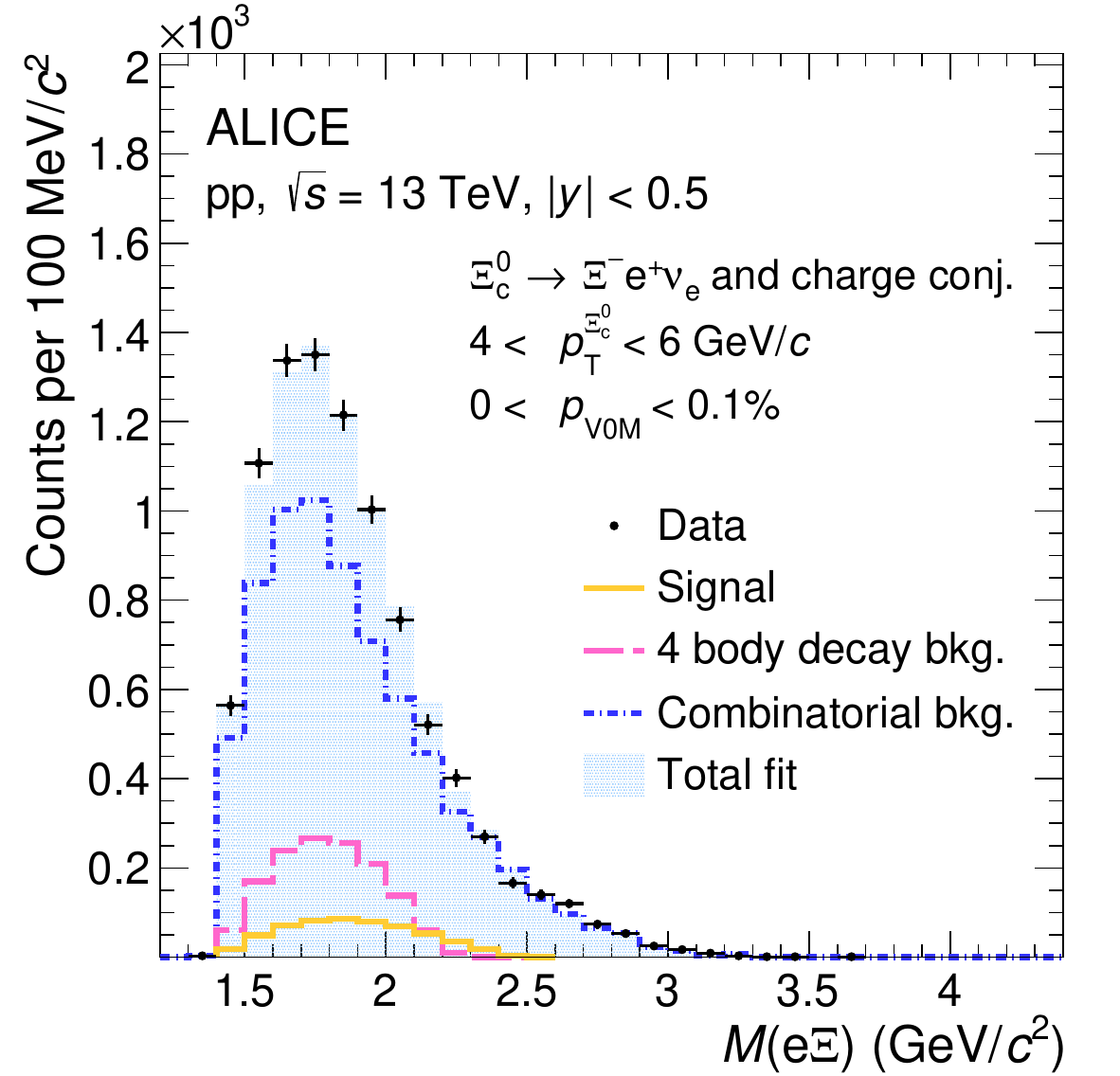}
    \includegraphics[width=0.475\textwidth]{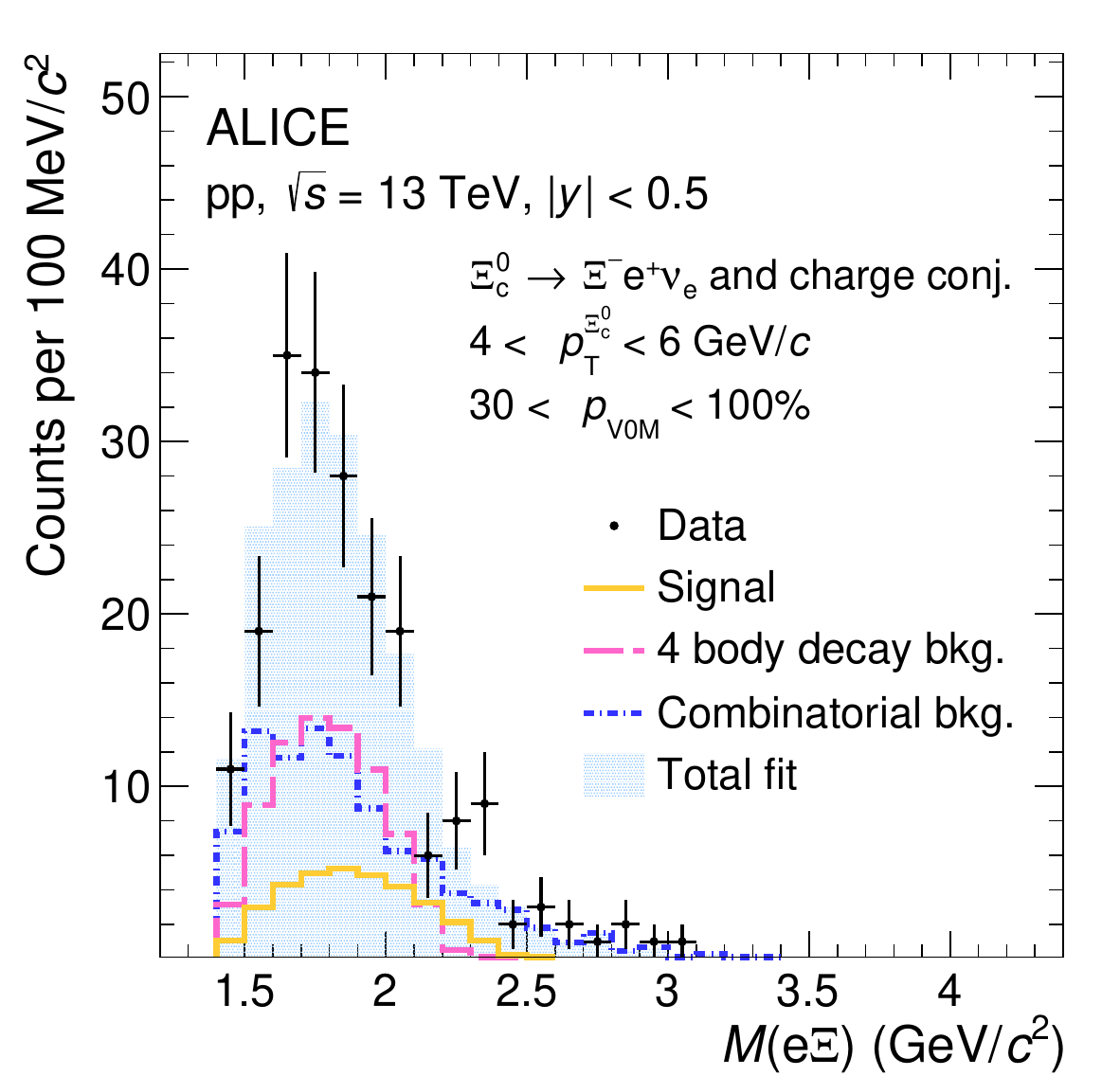}
    \caption{Invariant-mass distribution of \XiePair pairs for \XicZeroToXiEleNu in 4 $<$ \pt $<$ 6 \GeVc, in the high- (left) and low- (right) multiplicity classes. The blue filled distribution shows the total fit and the coloured lines indicate the different sources contributing to the fit.}
    \label{fig:Invamass_semilep}
\end{figure}

For the \Xiczero semileptonic decay, the \Xiczero invariant mass cannot be fully reconstructed due to the presence of a neutrino in the decay chain, and the signal candidates were obtained by pairing an electron and a $\Xi$ of opposite charges. Candidate electron tracks were selected by requiring at least three hits in the \ITS with two in the \SPD, a minimum of 50 reconstructed clusters and 70 crossed pad rows in the \TPC, \pt $>$ 0.5 \GeVc, and identified with the \TPC and \TOF detectors by requiring selection criteria 
$\left | n_{\sigma}^{\rm TPC} \right | < 3$ and
$\left | n_{\sigma}^{\rm TOF} \right | < 3$~\cite{ALICE:2024xjt}.
The requirement for hits in both layers of the \SPD was used to reject electrons from late photon conversions~\cite{ALICE:2018yau,ALICE:2019nuy}. Lastly, to suppress the electrons from Dalitz decays, each track was paired with opposite-sign tracks in the same event, and the track was discarded if it formed at least one pair with an invariant mass smaller than 50 \MeVmass~\cite{ALICE:2021psx}.
Once the \XiePair pairs were built, they were required to be within \yrangeAb{0.5} and have masses (${M_{\XiePair}}$) larger than 1.3 \GeVmass. In addition, the pairs were filtered out by a criterion based on the opening angle between the electron and the $\Xi$ momentum directions, which was obtained by studying the signal (\XicZeroToXiEleNu) in MC. The criterion was determined by the angle at which 90\% of \XiePair pair yields remain and was tuned for each \pt interval in order to minimise the rejection of signal candidates in the data. The resulting values were distributed from 60\degree to 23\degree, decreasing with increasing \pt.

The selected \XiePair pairs are composed of three different contributions:
i) the pairs from \XicZeroToXiEleNu (signal);
ii) the pairs from other decay channels
$\Xiczero \rightarrow \rm e^{+}\Xi^{\ast-}\nu_{\rm e}\xspace \rightarrow \rm e^{+}(\Xi^{-}\pizero )\nu_{\rm e}$ and
$\Xicplus \rightarrow \rm e^{+}\Xi^{\ast0}\nu_{\rm e}\xspace \rightarrow \rm e^{+}(\Xi^{-}\pip){\nu}_{\rm e}$,
hereafter denoted as \textit{4-body decay} background;
and iii) combinatorial background.
Since the difference between the \XicZeroToXiEleNu signal and 4-body decay background is only a soft pion from the resonance decay, the invariant-mass distribution of the 4-body decay background is slightly shifted towards lower masses with respect to the signal one, as shown in Fig.~\ref{fig:Invamass_semilep}.
To measure the reconstructed \XicZeroToXiEleNu yield, a corresponding template for each type of \XiePair pair was prepared and used to fit the overall \XiePair pair candidates.
The sources of the template for each type of \XiePair pair were as follows: i) \XicZeroToXiEleNu signal from MC; ii) 4-body decay background from MC; and iii) combinatorial background of \XiePair pairs modelled with same-charge sign pairs from data~\cite{ALICE:2021bli}.
Fig.~\ref{fig:Invamass_semilep} shows examples of the template fit for the \pt interval \ptrange{4}{6} \GeVc in the high- and low-multiplicity classes. In each panel, the distributions show the templates for each type of \XiePair pair, and the total fit indicates the sum of the templates after the fit was performed. The reliability of the template fit method was validated by an MC closure test, which will be explained in Section~\ref{sec:systErr}.
Once the \XiePair pairs from \XicZeroToXiEleNu were extracted, the \pt distribution of the pairs was corrected for the missing momentum carried by \nue, using a Bayesian unfolding technique~\cite{DAgostini:1994fjx} implemented in the \textrm{RooUnfold} package~\cite{Adye:2011gm}.


The \pt-differential production yield of prompt \Xiczeroplus per event in each multiplicity class was calculated as 

\begin{equation} 
   \frac{1}{N_{\mathrm{event}}}
   \frac{\mathrm{d}^2 N_{\mathrm{prompt}}^{\Xiczeroplus}}{\mathrm{d}p_{\mathrm{T}}\,\mathrm{d}y} =
   \frac{\epsilon_{\mathrm{trigger}}}{N_{\mathrm{event}}} \times
   \frac{1}{\mathrm{BR}} \times 
   \frac{1}{2 \Delta y \Delta p_{\mathrm{T}}}\times    
   \frac{f_{\mathrm{prompt}} \times 
   N_{\mathrm{raw}}^{{\Xiczeroplus}}}
   {(\mathrm{Acc} \times \epsilon)_{\mathrm{prompt}}}\,,
\label{eq:CrossSection2}
\end{equation}

where ${N_{\mathrm{event}}}$ denotes the number of events in the corresponding multiplicity class,
$\epsilon_{\mathrm{trigger}}$ is the trigger efficiency as reported in Table 2,
and BR represents the branching ratio of the considered decay channel.
The factor of 2 accounts for the average yield of particles and antiparticles. $\Delta y$ and $\Delta p_{\mathrm{T}}$ denote the rapidity coverage and the width of the \pt interval, respectively.
$f_{\mathrm{prompt}}$ refers to the fraction of prompt \Xiczeroplus in the extracted raw yield, $N_{\mathrm{raw}}^{{\Xiczeroplus}}$ is the raw yield (sum of particles and antiparticles),
and \mbox{$(\mathrm{Acc} \times \epsilon)_{\mathrm{prompt}}$} corresponds to the acceptance-times-efficiency for prompt \Xiczeroplus in the given multiplicity interval.

In the \acceff computation, the \pt distribution of \Xiczeroplus hadrons generated in MC simulations was tuned with an ad-hoc weight to better match the \pt distribution of the corrected data. To derive this weight, the ratio of the corrected yield of \Xiczeroplus in the analysed data to the generated \Xiczeroplus in MC was computed as a function of \pt. This \pt-dependent ratio was then fitted with an exponential function to obtain the weight.
An additional weight was assigned to account for the multiplicity dependence of the \acceff, which originates from the dependence of the primary-vertex resolution on the charged-particle multiplicity. The weight was defined as the ratio between the \ntrk distribution measured in data and MC, where \ntrk represents the number of track segments built by associating pairs of hits in the two layers of the SPD, which is proportional to the charged-particle multiplicity.
The \ntrk distributions from MC simulations were obtained by considering events with at least one \Xiczeroplus candidate, whose invariant mass lay within a specific range. For \Xiczeroplus hadronic decays, the range was based on the width of the signal peak obtained around the PDG mass~\cite{ParticleDataGroup:2022ynf}: $\pm20$ MeV from the central position for \XicZeroToPiXi and $\pm30$ MeV from the central position for \XicPlusToXiPiPi. For \Xiczero semileptonic decays, the range was ${M_{\XiePair}}$ $>$ 1.3 \GeV.
The weighted efficiencies computed for the different multiplicity classes show a maximum of 5\% variation with respect to those computed for the multiplicity-integrated sample for \Xiczero baryons. However, for \Xicplus, a maximum 15\% variation was observed in the high-multiplicity class, which decreased with increasing \pt.

The prompt fraction $f_{\mathrm{prompt}}$ in the total observed yield was estimated using a procedure similar to that described in Ref.~\cite{ALICE:2021bli}, to correct for the \Xiczeroplus yield originating from the beauty-hadron decays. In the estimation, it was assumed to be independent of the charged-particle multiplicity and therefore was obtained from the multiplicity-integrated event class (0 $<$ \pvom $<$ 100\%). This assumption is justified by the recent measurements of the non-prompt fraction of D mesons at midrapidity~\cite{ALICE:2023brx}, where the ratio of the non-prompt fraction in different multiplicity classes to that in the multiplicity-integrated class was found to be compatible with unity~\cite{ALICE:2023brx}.
The $f_{\mathrm{prompt}}$ was calculated as 
\begin{align}
    f_{\mathrm{prompt}} &= 1-\frac{N^{\Xiczeroplus}_{\text{non-prompt}}}{N^{\Xiczeroplus}_{\text{raw}}}\\ &=1-\frac{1}{N^{\Xiczeroplus}_{\text{raw}}} \times
    \left(\frac{\mathrm{d}^2\sigma}{\mathrm{d}\pt \mathrm{d}y}\right)^{{\Xiczeroplus}}_{\text{non-prompt}}
    \times {\rm BR} \times 2 {\Delta y} \Delta {\pt}
    \times {(\rm Acc\times \epsilon)_{\text{non-prompt}}} \times \mathcal{L}_{\mathrm{int}},
    \label{eq:feeddown}
\end{align}
where $N^{\Xiczeroplus}_{\text{raw}}$ and $N^{\Xiczeroplus}_{\text{non-prompt}}$ indicates the extracted raw yields and non-prompt yields,
$\left(\frac{\mathrm{d}^2\sigma}{\mathrm{d}\pt \mathrm{d}y}\right)^{{\Xiczeroplus}}_{\text{non-prompt}}$ denotes the cross section of the non-prompt \Xiczeroplus baryons,
BR is the branching ratio of the corresponding decay channel,
factor 2 accounts for the cross section being the average of particles and antiparticles, $\Delta y$ and $\Delta {\pt}$ represent the rapidity coverage and the width of the $\pt$ interval, respectively,
$(\rm{Acc} \times \epsilon)_\text{non-prompt}$ indicates the acceptance-times-efficiency for non-prompt \Xiczeroplus baryons,
and $\mathcal{L}_{\mathrm{int}}$ denotes the integrated luminosity.
Since the production cross section of non-prompt \Xiczeroplus baryons has not been measured yet, it was estimated by scaling the non-prompt \Lambdac cross section, under the assumption that the \pt-differential cross section of non-prompt \Xiczeroplus has a similar shape to that of non-prompt \Lambdac as 
\begin{equation}
    \left(\frac{\mathrm{d}^2\sigma}{\mathrm{d}\pt \mathrm{d}y}\right)^{{\Xiczeroplus}}_{\text{non-prompt}}
    \approx 
    \left(\frac{\mathrm{d}^2\sigma}{\mathrm{d}p_{\rm T}\mathrm{d}y}\right)^{\Lambdac}_{\text{non-prompt}} 
    \cdot
    \frac{ \sum_{h_\mathrm{b}} f(\mathrm{b}\rightarrow h_\mathrm{b} \rightarrow \Xi_\mathrm{c})}{\sum_{h_\mathrm{b}} f(\mathrm{b}\rightarrow h_\mathrm{b} \rightarrow \Lambda_\mathrm{c})}\, \\,
    \label{eq:PF1}
\end{equation}
where $h_\mathrm{b}$ indicates a beauty-hadron species and the sum runs on all beauty-hadron species.
The cross section of non-prompt \Lambdac was estimated by combining the b-quark production cross section from FONLL calculations~\cite{Cacciari:2012ny} with the fragmentation fraction of b-quarks decaying into beauty hadrons measured by LHCb
experiment~\cite{LHCb:2019fns}. The kinematics of beauty hadrons decaying into \Lambdac were simulated using the PYTHIA 8 event generator. The obtained non-prompt \Lambdac cross section was scaled with the rightmost term of Eq.~\ref{eq:PF1}, which represents the fragmentation fraction of non-prompt \Xiczeroplus to that of non-prompt \Lambdac, in order to estimate the non-prompt \Xiczeroplus cross section.
By considering that only $\Lambda_{\mathrm{b}}^0$ and $\Xi_{\mathrm{b}}^-$ baryons contribute to the yields of non-prompt \Lambdac and \Xiczeroplus baryons, respectively, the ratio was approximated as Eq.~\ref{eq:PF2}.
It was estimated by scaling the measured prompt \Xiczeroplus/\Lambdac cross section ratio~\cite{ALICE:2021rzj,ALICE:2021bli} with the production yield ratio of non-prompt to prompt \Xiczeroplus/\Lambdac predicted by PYTHIA 8 with CR Mode 2 tune. This approach relies on the assumption that the \pt-differential cross section ratio of non-prompt \Xiczeroplus/\Lambdac has a similar shape to that of the prompt case.
\begin{equation}
    \frac{ \sum_{h_\mathrm{b}} f(\mathrm{b}\rightarrow h_\mathrm{\mathrm{b}} \rightarrow \Xi_\mathrm{c})}{\sum_{h_\mathrm{b}} f(\mathrm{b}\rightarrow h_\mathrm{b} \rightarrow \Lambda_\mathrm{c})}
    = 
    \frac{f(\mathrm{b}\rightarrow\Xi_\mathrm{b}\rightarrow \Xi_\mathrm{c})}{f(\mathrm{b}\rightarrow\Lambda_\mathrm{b} \rightarrow \Lambda_\mathrm{c})}\\
    =
    \left(\frac{\frac{\mathrm{b}\rightarrow\Xi_\mathrm{c}}{\mathrm{b}\rightarrow \Lambda_\mathrm{c}}}{\frac{\mathrm{c}\rightarrow\Xi_\mathrm{c}}{\mathrm{c}\rightarrow \Lambda_\mathrm{c}}}\right)_{\mathrm{PYTHIA~8}}
    \cdot\frac{({\mathrm{d}^2\sigma}/{\mathrm{d}\pt \mathrm{d}y})^{\mathrm{\Xiczeroplus}}_{   
    \mathrm{prompt}}}{({\mathrm{d}^2\sigma}/{\mathrm{d}\pt \mathrm{d}y})^{\mathrm{\Lambdac}}_{\mathrm{prompt}}}
    \label{eq:PF2}
\end{equation}
The estimated prompt fraction ranges between 0.94 and 0.98.

\section{Systematic uncertainties}\label{sec:systErr}

Systematic uncertainties on the prompt \Xiczeroplus yield measurements were estimated for each \pt interval and multiplicity class.
The sources of systematic uncertainty were those relative to the raw yield extraction, the MC \pt shape, the computation of the prompt fraction, the efficiency correction, the tracking efficiencies, the unfolding procedure, and the $\left | y \right |$ variation.
The uncertainties estimated in the multiplicity-integrated event class (0 $<$ \pvom $<$ 100\%) were reported in Table~\ref{tab:systErr}, while those for other multiplicity classes were provided in Appendix~\ref{app:systerr}.
The magnitudes of the uncertainty contributions were independent of multiplicity. The tracking efficiency, the PID selection related to the efficiency correction, and the generated \Xiczeroplus \pt shape were correlated across different multiplicity intervals, while the others were not.
The contribution related to the PID selection was included in the systematic uncertainty associated with the efficiency correction.

\renewcommand{\arraystretch}{1.35}
\begin{table}[t]
\caption{Systematic uncertainties on the corrected prompt yield for the multiplicity-integrated event class}
\centering
\resizebox{\textwidth}{!}
{
    \begin{tabular}{c|rrr|rrrr|rrrr}
    \hline\hline
    Decay channel & \multicolumn{3}{c|}{\XicPlusToXiPiPi} & \multicolumn{4}{c|}{\XicZeroToPiXi} &
    \multicolumn{4}{c}{\XicZeroToXiEleNu} \\
    \pt (\GeVc) & 4--6 & 6--8 & 8--12    & 2--4 & 4--6 & 6--8 & 8--12    & 2--4 & 4--6 & 6--8 & 8--12
    \\\hline
    Raw yield       & 5\% & 5\% & 6\%    & 6\% & 7\% & 8\% & 9\%    & 21\% & 11\% & 10\% & 5\% \\
    MC \pt shape    & 1\% & negl. & negl.    & 3\% & 1\% & negl. & negl.    & 5\% & 2\% & 1\% & 2\% \\    
    MC multiplicity & negl. & negl. & negl.    & negl. & negl. & negl. & negl.    & 1\% & 1\% & 1\% & 2\% \\
    Prompt fraction & $^{+4}_{-6}$\% & $^{+4}_{-6}$\% & $^{+6}_{-7}$\%
                    & $^{+3}_{-2}$\% & $^{+3}_{-2}$\% & $^{+4}_{-3}$\% & $^{+4}_{-3}$\%
                    & $^{+5}_{-1}$\% & $^{+6}_{-1}$\% & $^{+4}_{-1}$\% & $^{+4}_{-3}$\% \\
    Efficiency correction & 9\% & 11\% & 11\%   & 6\% & 6\% & 6\% & 5\%   & 7\% & 5\% & 6\% & 6\% \\
    Tracking   & 5\% & 5\% & 5\%    & 5\% & 5\% & 5\% & 5\%    & 5\% & 5\% & 5\% & 5\% \\
    Unfolding  & -- & -- & --       & -- & -- & -- & --        & 1\% & 4\% & 3\% & 4\% \\
    $\left | y \right |$ variation & -- & -- & --  & -- & -- & -- & --  & 1\% & 1\% & 1\% & 1\% \\\hline
    Total & 13\% & 14\% & 15\%    & 11\% & 10\% & 12\% & 12\%   & 23\% & 15\% & 14\% & 12\% \\
    \hline
    Branching ratio & \multicolumn{3}{c|}{44.3\%} & \multicolumn{4}{c|}{18.9\%} & \multicolumn{4}{c}{23.1\%} \\
    \hline\hline
    \end{tabular}
}
\label{tab:systErr}
\end{table}

The approach adopted to estimate the systematic uncertainty related to the raw-yield extraction depends on the \Xiczeroplus decay channel. For the \XicPlusToXiPiPi and \XicZeroToPiXi decays, the uncertainty was evaluated by repeating the fitting procedure to the invariant-mass distributions in each \pt and multiplicity class with various background modelling functions and fitting ranges. 
In addition, a bin-counting method was employed to test the sensitivity to the signal shape by following the approach used in Ref.~\cite{ALICE:2020wfu}. The signal function width was fixed to the value observed in MC simulations and a variation of 10\% was also applied. The root-mean-square (RMS) of the raw yield distribution and the deviation of the mean of the distribution with respect to the raw yield obtained for the central case were evaluated, and their quadratic sum was assigned as a systematic uncertainty. The final assigned uncertainties range from 5\% to 6\% for \Xicplus and from 6\% to 9\% for \Xiczero, depending on the \pt of the \Xiczeroplus baryon.
For the \XicZeroToXiEleNu decay, the uncertainty was evaluated using an MC closure test. To test the reliability of the template fit method discussed in Section~\ref{sec:dataAna}, multiple pseudo-data samples with a known (true) signal yield were created, and then template fits were performed on these pseudo-data samples to check the difference between the true and the measured signal yield. To create a sample, each type of \XiePair pair was randomly sampled from its base template, with the same number of candidates as in the data in each \pt interval, with the signal fraction varying from 5\% to 25\%. The systematic uncertainty was estimated as the deviation between the true signal yield and the measured signal yield $(1 - N_{\rm measured}/{N_{\rm true}})$. The final assigned uncertainties range from 5\% to 21\%.

To estimate the uncertainty associated with the \pt shape of the generated \Xiczeroplus baryons in MC simulations, the weighting factor described in Section~\ref{sec:dataAna} was varied within the uncertainties of the ratio between the data and the MC. The maximum assigned systematic uncertainty was 5\%.

In the MC, the multiplicity was estimated with the \ntrk distribution, which was corrected using weights. The systematic uncertainties of these weights were calculated independently among the decay channels. For the \Xiczeroplus hadronic decay channels, the event selection criteria for determining the weights were tested by varying the invariant-mass window of \Xiczeroplus candidates with respect to the PDG mass~\cite{ParticleDataGroup:2022ynf} and by removing the mass window requirement.
The resulting uncertainty was negligible except for \Xiczero baryons with \pt $<$ 6 \GeVc in the high-multiplicity class, where the uncertainty was found to be at maximum of 4\%.
For the semileptonic decay of \Xiczero, the weights were obtained from the events with \Xiczero candidates (\XiePair pairs) having a mass within the range {\mbox{$1.3 < M_{\XiePair} < 2.5$}}~\GeVmass. The uncertainty was 1-2\% for most cases except in the high-multiplicity class, where the maximum assigned systematic uncertainty was 5\%.

To estimate the uncertainty in the prompt \Xiczeroplus fraction correction described in Eqs.~\ref{eq:PF1} and~\ref{eq:PF2}, the following two ingredients were considered: i) the non-prompt \Lambdac cross section from FONLL predictions and ii) the ratio of the fractions of b quarks contributing to \Xiczeroplus and \Lambdac yields
$(f(\mathrm{b} \rightarrow \Xi_\mathrm{b} \rightarrow \Xi_\mathrm{c})$ /
$f(\mathrm{b} \rightarrow \Lambda_\mathrm{b} \rightarrow \Lambda_\mathrm{c}))$,
as done in Ref.~\cite{ALICE:2021bli}.
The uncertainties of the FONLL prediction were estimated by varying the b-quark mass, factorisation scale, and renormalisation scale as prescribed in Ref.~\cite{Cacciari:2012ny}.
The uncertainty on the b-quark fractions feeding down to \Xiczeroplus and \Lambdac baryons was estimated by setting upper and lower bounds as follows. For the upper bound, only the measured prompt \Xiczeroplus/\Lambdac production yield ratios were considered, without scaling the ratio of non-prompt to prompt \Lambdac and \Xiczeroplus yields predicted by PYTHIA 8, to account for the possible differences between the $\Xiczeroplus/\Lambdac$ and $\Xib/\Lambdab$ ratios. For the lower bound, the \Xib/\Lambdab ratio measured by the LHCb Collaboration~\cite{LHCb:2019fns} was considered. Here it was assumed that the relative contribution of beauty-hadron decays to \Xiczero and \Xicplus in different multiplicity classes remains constant. The maximum resulting uncertainty was 4\%.
To estimate potential deviations of  $f_{\rm prompt}$ from the value computed in the multiplicity-integrated event class, a similar methodology to that in Ref.~\cite{ALICE:2021npz} was utilised. The variation as a function of multiplicity was computed using PYTHIA~8 simulations. The systematic uncertainty was evaluated considering the ratio between the non-prompt fraction in a given multiplicity class and that in INEL $>$ 0 events. The maximum uncertainty was 3\%. 
The uncertainty of the prompt fraction was calculated by taking the quadratic sum of two sources. The maximum uncertainty was 7\%.

The uncertainty related to the efficiency correction associated with the selection of the BDT classification score was estimated by repeating the analysis for different threshold values on the BDT scores to obtain the prompt \Xiczeroplus yield. The uncertainty was estimated as the quadratic sum of the RMS of the corrected prompt \Xiczeroplus yield distribution obtained from the variations and the shift in its mean with respect to the yield obtained with the default threshold value.
These variations in threshold values were limited to trials with a \Xiczeroplus signal that has a statistical significance of more than 3$\sigma$ to be less sensitive to statistical fluctuations.
The resulting uncertainties ranged from 9\% to 11\% for the \XicPlusToXiPiPi decay, and from 5\% to 6\% for the \XicZeroToPiXi decay. 
For the \XicZeroToXiEleNu decays, a set of variations of the selection criteria was considered for each variable, and the maximum deviation from the result obtained with respect to the default selection values was assigned as the uncertainty. The procedure was repeated for all selection variables and the final uncertainty was calculated as the quadratic sum of the uncertainties associated with each variable. The estimated uncertainties from these variations ranged from 5\% to 7\%.

The systematic uncertainty related to the track reconstruction efficiency can be influenced by the criteria related to the track quality selection and the probability of prolonging tracks from the \TPC to the \ITS (matching efficiency). The effect of the first source was evaluated by calculating the prompt yields of \Xiczeroplus baryons with varied track selection criteria. The RMS of the calculated prompt yields was assigned as the systematic uncertainty. As a result, a uniform 5\% uncertainty was assigned for all \pt intervals of interest. To estimate the contribution from the second source, the uncertainty in the matching efficiency of the pion (for hadronic \Xiczeroplus decay channels) and the electron (for the semileptonic \Xiczero decay channel) in data and MC were compared. The per-track uncertainty on the matching efficiency was propagated to \Xiczeroplus candidates by taking the decay kinematics into account. In this evaluation, only matching efficiencies of pions and electrons were considered since the prolongation of the tracks from the \TPC to the \ITS hits was not required for the tracks originating from the $\Xi$ decay.
The resulting uncertainty ranged from 1\% to 2\%.

For the semileptonic decay channel of \Xiczero, two additional sources were considered for the total systematic uncertainty. The first source was the unfolding procedure to compensate for the missing momentum carried by \nue. To estimate the uncertainty of the procedure, both the algorithm (Bayesian and Singular Value Decomposition~\cite{Hocker:1995kb}) and the number of iterations (2$-$6) were varied. The estimated uncertainties ranged from 1\% to 4\%. The second source came from possible differences in the acceptance of \XiePair pairs between the data and the MC. To estimate the uncertainty, the rapidity interval was varied between \yrangeAb{0.5} and 0.8. The estimated uncertainties were uniform for all \pt intervals at 1\%.

The systematic uncertainties described above are assumed to be uncorrelated with one another. Therefore, the total systematic uncertainty was calculated as the quadratic sum of each source for each \pt and multiplicity class.

\section{Results}\label{sec:result}

\begin{figure}[b]
    \centering
    \includegraphics[width=0.95\textwidth]{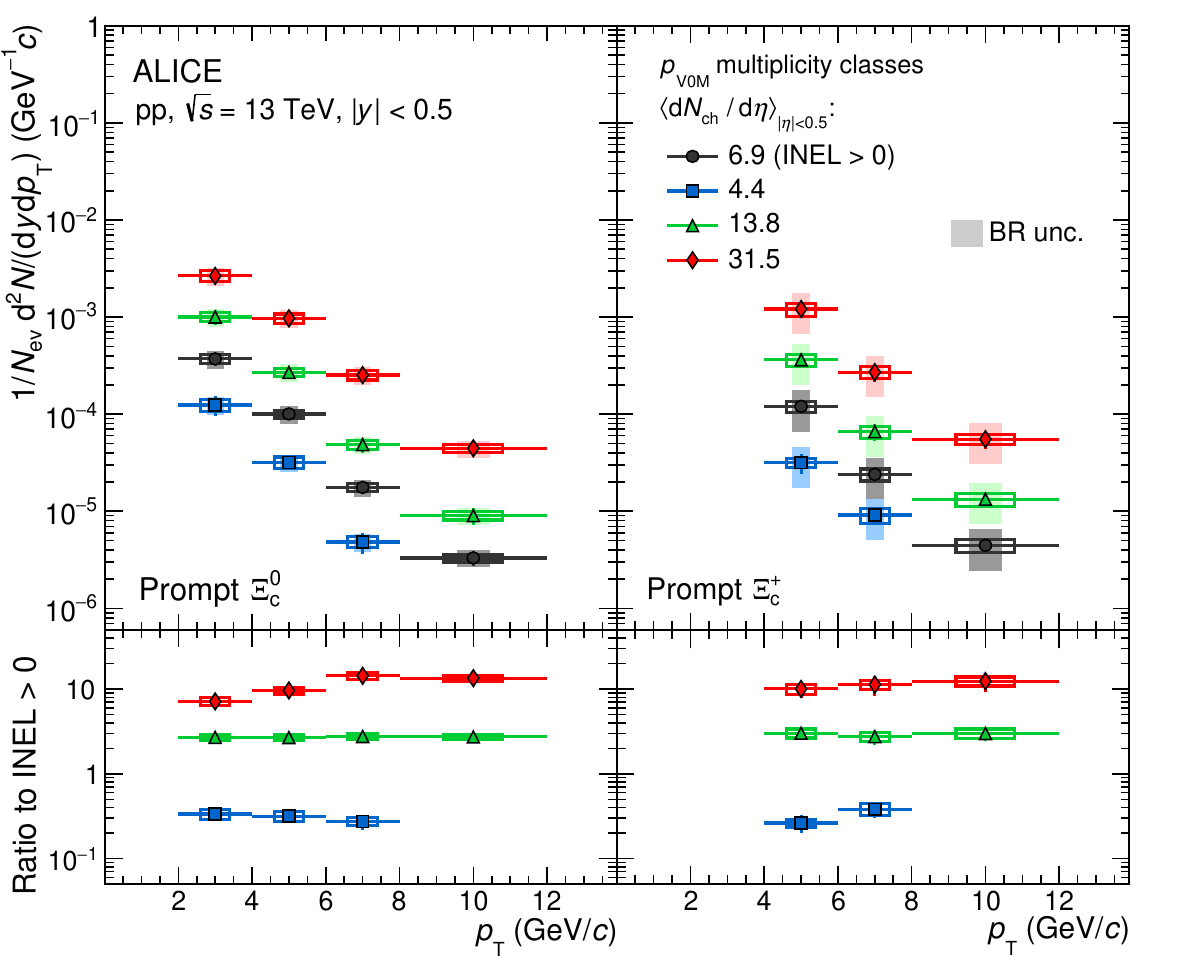}
    \caption{\pt-differential per-event yield of prompt \Xiczero (left) and \Xicplus (right) baryons measured in the different multiplicity classes in pp collisions at $\sqrt{s}=13$ TeV at midrapidity ($|y| < 0.5$), along with the corresponding ratios to the multiplicity-integrated (INEL $>$ 0) class in the bottom panel. The values shown in the legend, denoted as $\left< \dndeta \right>_{|\eta|<0.5}$, correspond to the average charged-particle multiplicity at midrapidity for the respective multiplicity classes, as introduced in Table~\ref{tab:multiplicity class}. The statistical and systematic uncertainties are shown as bars and open boxes, respectively. The shaded boxes indicate the uncertainty of the branching ratio.}
    \label{fig:CorrectedYieldPerEvent}
\end{figure}
%
The corrected prompt yields of the \Xiczero from both hadronic and semileptonic decay channels are consistent within the statistical and systematic uncertainties uncorrelated over the particle types. The following systematic uncertainty sources are considered uncorrelated: raw yield, MC \pt shape, efficiency correction, unfolding, $\left | y \right |$ variation, and branching ratio.
To obtain a result with better precision, a weighted average of the two measurements was computed with weights defined as the inverse of the quadratic sum of the relative statistical and uncorrelated systematic uncertainties.
Figure~\ref{fig:CorrectedYieldPerEvent} shows the \pt-differential yields of prompt \Xiczero and \Xicplus baryons measured in \yrangeAb{0.5} in the multiplicity-integrated class (INEL $>$ 0) and in three different charged-particle multiplicity classes in \pp collisions at \thirteen.
The statistical and systematic uncertainties are reported as vertical bars and open boxes, respectively, and the uncertainties from the branching ratio are displayed as shaded boxes.
The multiplicity classes are represented in terms of the average charged-particle densities at midrapidity, $\left< \dndeta \right>_{|\eta|<0.5}$ values, as reported in Table ~\ref{tab:multiplicity class}.
The top left panel of Fig.~\ref{fig:CorrectedYieldPerEvent} shows the average result between the \Xiczero measurements in the hadronic and semileptonic decay channels measured in \ptrange{2}{12} \GeVc and the top right panel shows \Xicplus measured in \ptrange{4}{12} \GeVc. The prompt \Xiczero yields measured in four multiplicity classes are compatible with those of \Xicplus baryons, as expected from isospin symmetry, and as was observed in the previous measurement~\cite{ALICE:2021bli}.
%
The bottom panels of Fig.~\ref{fig:CorrectedYieldPerEvent} show the ratios between the \Xiczeroplus yield in a given multiplicity class and that obtained in INEL $>$ 0 events.
To calculate the ratio, the correlation of the uncertainty sources between the multiplicity classes and the INEL $>$ 0 class was considered as follows: i) the high-multiplicity class trigger (HMV0) and MC multiplicity were treated as uncorrelated; ii) raw yield extraction and cut variations were assumed to be partially correlated, where the largest uncertainty was chosen as the final contribution to the total uncertainty; and iii) the other sources were considered as fully correlated.
The observed hardening trend of the \pt spectrum is compatible with that reported for D mesons and \Lambdac in Ref.~\cite{ALICE:2021npz}. However, the current measurement precision does not allow one to exclude a smoother, or even negligible, evolution of the \pt distribution with multiplicity.


\begin{figure}[t]
\centering
\includegraphics[width=0.95\textwidth]{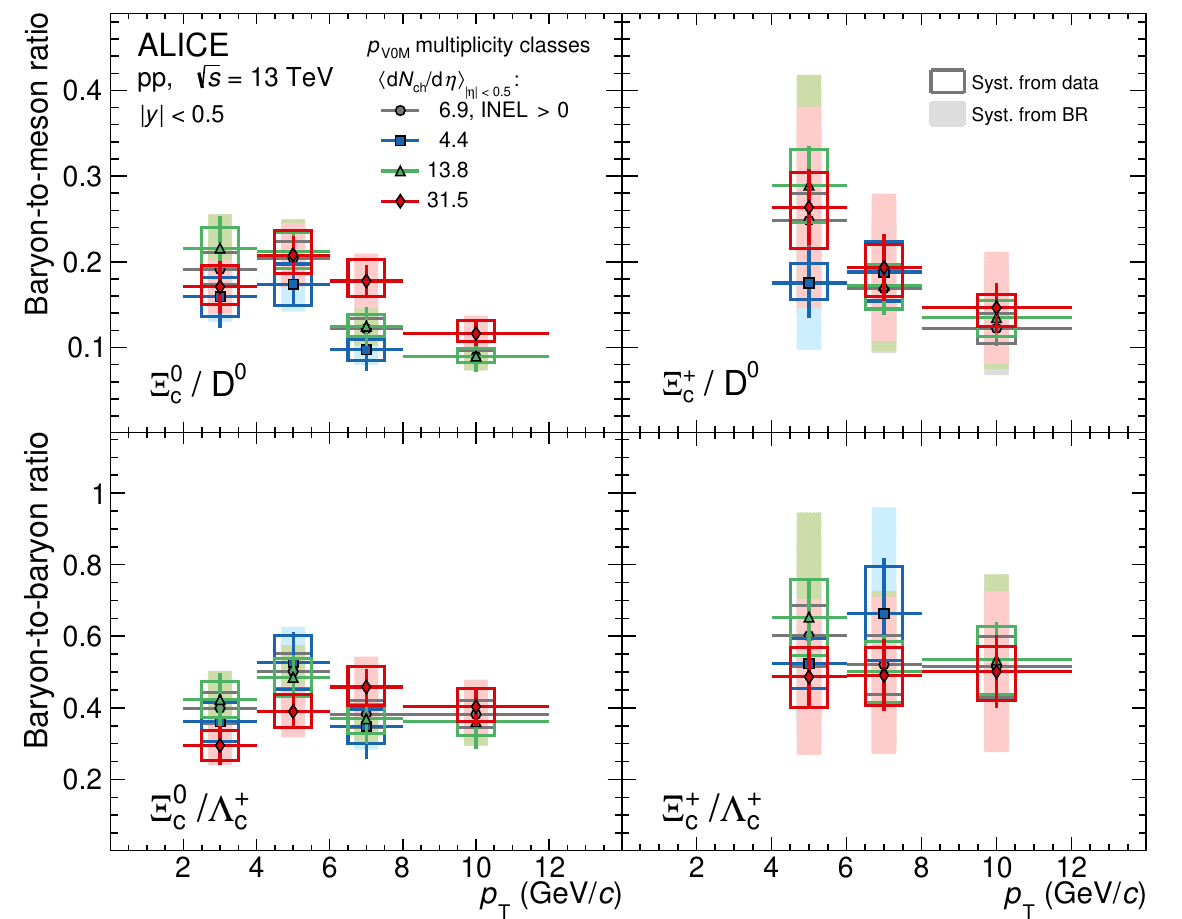}
\caption{The prompt production yield ratios between \Xiczeroplus and \Dzero mesons (top) and \Xiczeroplus and \Lambdac baryons (bottom) measured in the same multiplicity classes in pp collisions at $\sqrt{s}=13$ TeV~\cite{ALICE:2021npz}. The statistical and systematic uncertainties are shown as bars and open boxes, and the uncertainty from BR is represented in shaded boxes, respectively.}
\label{fig:XicOverDandXicOverLc}
\end{figure}

Figure~\ref{fig:XicOverDandXicOverLc} shows the baryon-to-meson ratios \Xiczeroplus/\Dzero and the baryon-to-baryon ratios \Xiczeroplus/\Lambdac in the measured multiplicity classes. In each panel, the bars (open boxes) indicate the statistical (systematic) uncertainty, while the shaded boxes represent the branching ratio uncertainties. For the propagation of the systematic uncertainty, the following uncertainty sources were treated as uncorrelated over the different baryon and meson measurements: raw yield extraction, MC \pt shape, efficiency correction, unfolding, $\left | y \right |$ variation, and branching ratio. Both the \Xiczeroplus/\Dzero and \Xiczeroplus/\Lambdac ratios measured in the INEL $>$ 0 class are consistent with the measurements in Ref.~\cite{ALICE:2021bli} within the uncertainties.
%
The measured \Xiczeroplus/\Dzero ratios show no significant dependence on \pt within the current experimental uncertainties. The ratios measured in high-multiplicity classes are compatible with those measured in low-multiplicity classes.
In contrast, the \pt-dependent \Lambdac/\Dzero ratio exhibits an increasing trend with charged-particle multiplicity, indicating a possible modification of the hadronisation process in high-multiplicity environments~\cite{ALICE:2021npz}.
%
Interestingly, however, the \pt-integrated \Lambdac/\Dzero ratio shows no significant dependence on multiplicity~\cite{ALICE:2021npz}, similarly to the behaviour observed for the $\Lambda$/$\mathrm{K}^0_{\mathrm{S}}$ ratio, while the \pt-integrated $\X$/$\mathrm{K}^0_{\mathrm{S}}$ and $\Om$/$\mathrm{K}^0_{\mathrm{S}}$ ratios increase with
increasing charged-particle multiplicity~\cite{ALICE:2019avo}.


\begin{figure}[b]
    \centering
    \includegraphics[width=0.95\textwidth]{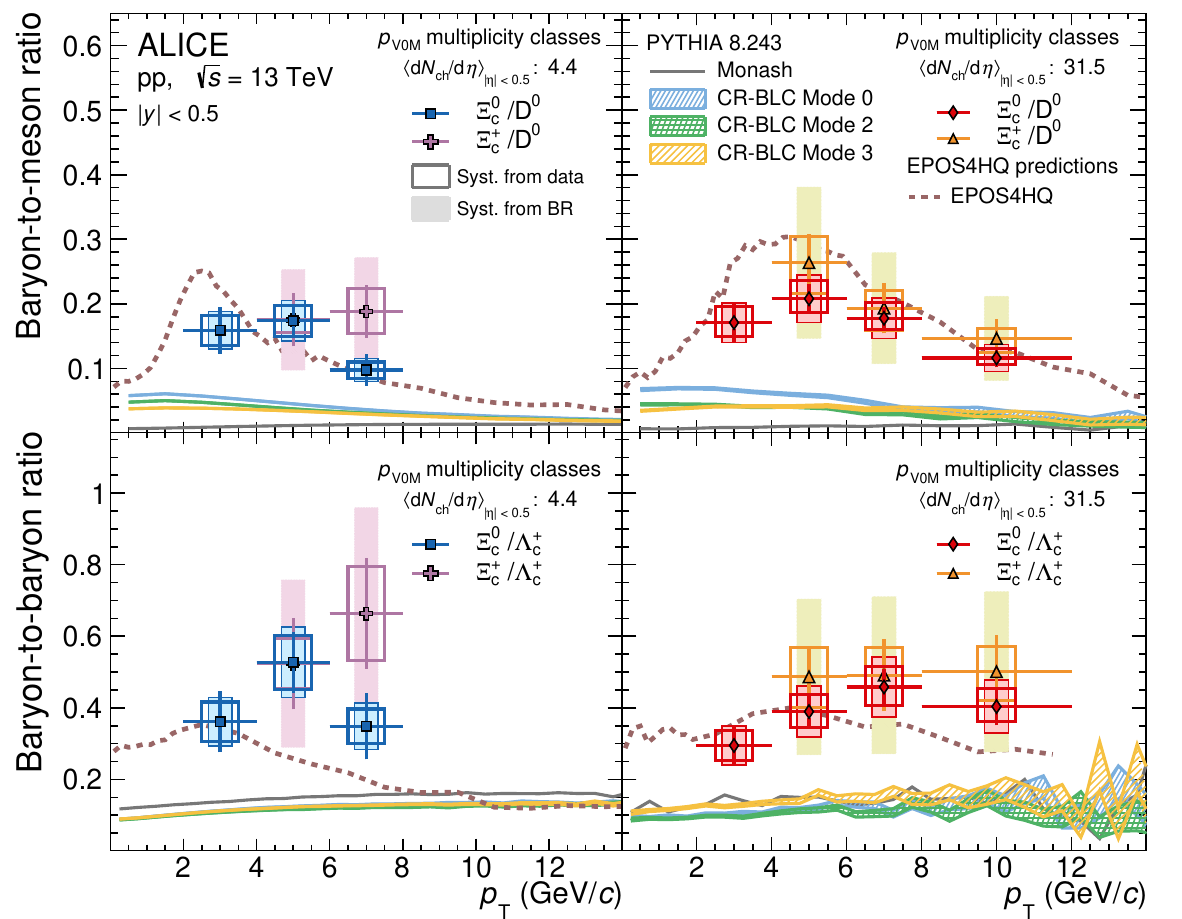}
    \caption{The baryon-to-meson ratios (top) and baryon-to-baryon ratios (bottom), measured in the low- (left) and high- (right) multiplicity classes. The measurements are compared with the predictions from two event generators: PYTHIA 8 with different tunes (namely Monash~\cite{Skands_2014}, CR-BLC~\cite{Christiansen:2015yqa} Mode 0, 2, and 3) and EPOS4HQ~\cite{EPOS4HQ}.}
    \label{fig:CharmedHadronRatiosTheory}
\end{figure}

Figure~\ref{fig:CharmedHadronRatiosTheory} shows the baryon-to-meson ratios \Xiczeroplus/\Dzero and baryon-to-baryon ratios \Xiczeroplus/\Lambdac obtained from the low- and the high-multiplicity classes, compared with model calculations from the PYTHIA~8.2~\cite{Sjostrand:2014zea} and the EPOS4HQ~\cite{EPOS4HQ} event generators.
Both the PYTHIA~8.2 and the EPOS4HQ predictions use the same multiplicity classes as the measurements. Note that both predictions are computed by considering \Xiczero only, however, the predictions for \Xicplus won't be significantly different being the isospin partner of \Xiczero.
The PYTHIA~8.2 simulations are obtained by using the standard Monash 2013 tune and the colour reconnection settings beyond the leading colour approximation (CR-BLC)~\cite{Christiansen:2015yqa}. The CR-BLC modes used in this study (0, 2, and 3) apply different constraints on the allowed reconnection among colour sources, which leads to increased baryon production.
In the EPOS4HQ, after the initial parallel scatterings, the medium is separated into core and corona components, and then the evolution of the core is modelled using viscous hydrodynamics. Heavy quarks are produced initially via time-like cascades, space-like cascades, and Born processes. A heavy quark may enter the fluid, and propagate through the medium, interacting with thermal partons through elastic and inelastic scatterings, during which it can lose or gain energy. When the local energy density drops below a critical threshold, the heavy quark may hadronise into various heavy-flavour hadrons via a coalescence mechanism. In contrast, heavy quarks which do not enter the fluid hadronise always via fragmentation. 
The measured \pt-differential baryon-to-meson ratios and baryon-to-baryon ratios show no significant dependence on multiplicity with the current uncertainties.
The predictions from PYTHIA~8.2 show no clear trend with multiplicity for Monash and CR-BLC mode 2. However, a slight increase can be observed for CR-BLC modes 0 and 3 in high multiplicity at low \pt. For both \Xiczeroplus/\Dzero and \Xiczeroplus/\Lambdac, both the Monash tune and the CR-BLC modes substantially underestimate the measured ratios in all multiplicity classes.
On the other hand, the predictions from the EPOS4HQ show a relatively good description of the data. The predictions agree with the data both qualitatively and quantitatively considering the systematic uncertainty, especially at high multiplicity. It is also worth noting that the EPOS4HQ predictions show an evolution with multiplicity.
The main reason for this evolution is the fraction of charm quarks entering the fluid, since a fluid is created more frequently at high multiplicity and therefore has the chance to hadronise via coalescence, rather than via fragmentation. The latter strongly suppresses high masses compared to light ones, whereas the former only requires the presence of light quarks in the fluid to combine into charmed hadrons.
The measurements of \Xiczeroplus production as a function of charged-particle multiplicity can put further constraints on the description of hadronisation mechanisms into baryons with charm and strange valence quarks.
Aside from the \Xiczero/\Dzero ratio in this analysis, note that the PYTHIA~8 CR-BLC modes have shown a significant multiplicity dependence in the \Lambdac/\Dzero ratio~\cite{ALICE:2021npz}, especially for modes 0 and 2. In addition, they qualitatively describe the measured \pt-differential \Lambdac/\Dzero ratio, including the decreasing trend with \pt and the overall magnitude as a function of charged-particle multiplicity. Also, it is noteworthy that the PYTHIA~8 CR-BLC modes predict a dependence on the charged-particle multiplicity of the \pt-integrated \Lambdac/\Dzero ratio, which is not suggested by the experimental data~\cite{ALICE:2021npz}.


\begin{figure}[b]
    \centering
    \includegraphics[width=0.6\textwidth]{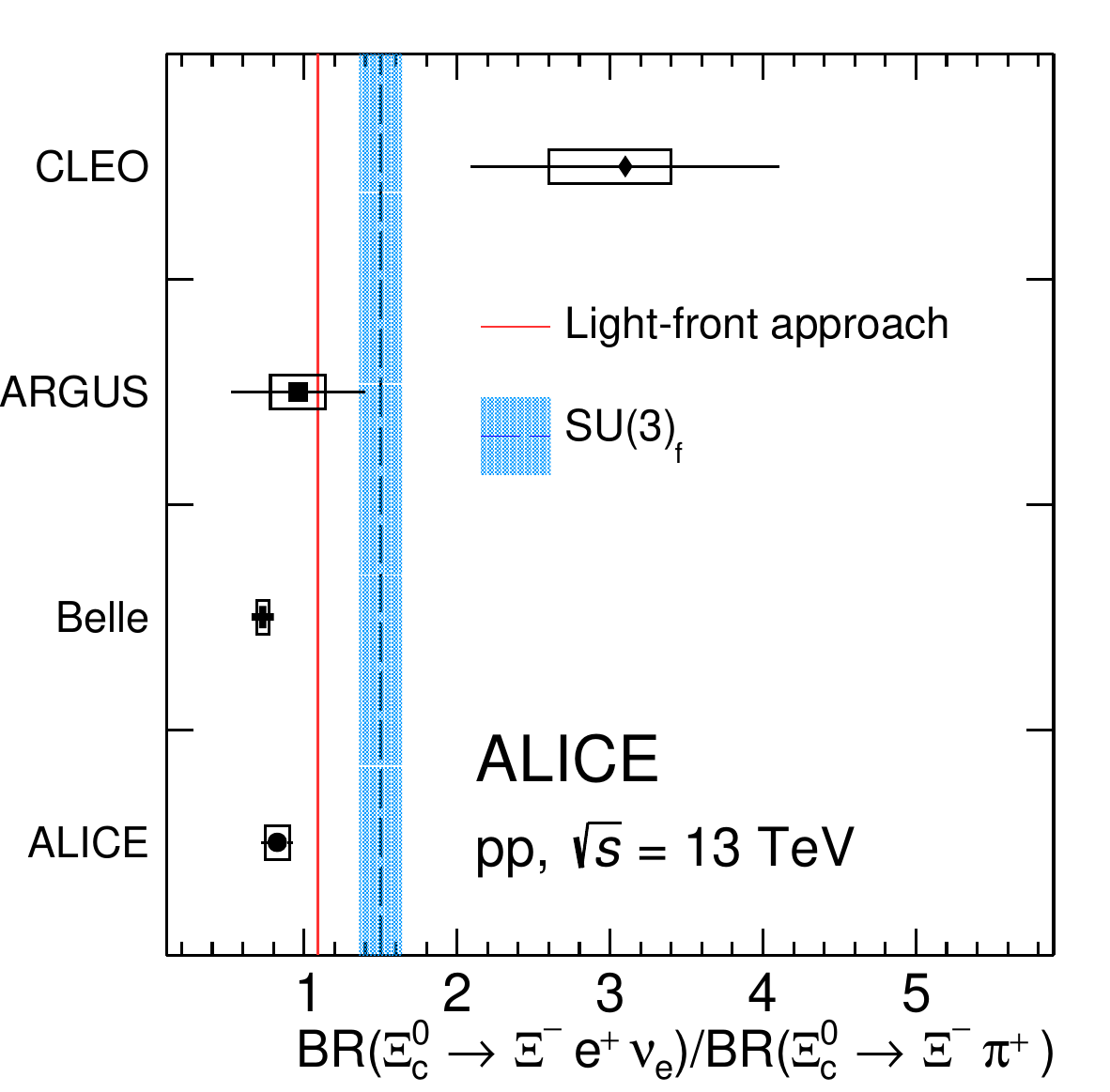}
    \caption{Comparison of BR(\XicZeroToXiEleNu)/BR(\XicZeroToPiXi) between experiments~\cite{ARGUS:1992jnv,Xic0_Belle,Xic0_CLEO} and model predictions~\cite{Zhao_2018, GENG2019214}}
    \label{fig:BRFrac}
\end{figure}

Figure~\ref{fig:BRFrac} shows the branching fraction ratio between the two considered decay channels of the \Xiczero baryon BR(\XicZeroToXiEleNu)/BR(\XicZeroToPiXi), compared with results from the ARGUS, Belle, and CLEO experiments~\cite{ARGUS:1992jnv, Xic0_Belle, Xic0_CLEO} and theoretical predictions~\cite{Zhao_2018, GENG2019214}. To calculate the branching fraction ratio, the production yield ratio between the two \Xiczero decay channels is obtained considering the \acceff corrected \Xiczero yields in the two decay channels without correcting them by the respective branching ratios. The measured ratio does not show any significant \pt dependence.
To provide a single value of the branching fraction ratio, the ratios in different \pt intervals were averaged by using the inverse of the sum in quadrature of the relative statistical and uncorrelated systematic uncertainties as weights as described in Refs.~\cite{ALICE:2021bli, ALICE:2024xjt}.
In the calculation, the following uncertainty sources are considered as \pt uncorrelated: the raw yield extraction (for both \XicZeroToPiXi and \XicZeroToXiEleNu), the BDT efficiency correction for the hadronic decay channel, and the unfolding procedure for the semileptonic decay channel.
The resulting branching fraction ratio is BR(\XicZeroToXiEleNu)/BR(\XicZeroToPiXi) =
0.825 $\pm$ 0.094 (stat.) $\pm$ 0.081 (syst.). This result supersedes the previous ALICE measurement~\cite{ALICE:2021bli} with a factor 1.6 improvement in statistical precision, due to the inclusion of both MB and HMV0 triggered data samples.
The branching fraction ratio is consistent with the Belle result~\cite{Xic0_Belle} within 0.72$\sigma$. However, the values predicted by theoretical models such as the Light-front approach~\cite{Zhao_2018} and SU(3)$\rm_f$~\cite{GENG2019214} overestimate the measured ratio, as illustrated in Fig.~\ref{fig:BRFrac}.
Also, it is an interesting point that the measured branching fraction ratio of \Xiczero is similar to that of \Omegac~\cite{ALICE:2024xjt} within 0.80$\sigma$, as predicted by model calculations using the Light-front approach and SU(3)f.


\section{Summary}\label{sec:summary}

The \pt-differential production yields of prompt \Xicplus and \Xiczero baryons at midrapidity (\yrangeAb{0.5}) as a function of charged-particle multiplicity, in the \pt interval \ptrange{4}{12} \GeVc for \Xicplus and \ptrange{2}{12} \GeVc for \Xiczero, are reported. This study presents the first multiplicity-dependent measurement of \Xiczeroplus baryons in \pp collisions at midrapidity.
The production yields for the minimum-bias multiplicity class INEL $>$ 0 are similar to the previously published result~\cite{ALICE:2021bli}.
No significant multiplicity dependence is observed in either \Xiczeroplus/\Dzero or \Xiczeroplus/\Lambdac within uncertainties. The EPOS4HQ prediction accurately describes the \Xiczeroplus/\Dzero ratio as a function of \pt and multiplicity, while the PYTHIA 8.2 Monash 2013 tune and CR-BLC modes fail to capture any feature of the measurement.
The branching fraction ratio BR(\XicZeroToXiEleNu)/BR(\XicZeroToPiXi) is also measured, and it supersedes the previous measurement~\cite{ALICE:2021bli} with a factor of 1.6 improvement in the statistical precision. This new branching fraction ratio is compatible with the Belle result within 0.72$\sigma$.
Lastly, the current measurement can be improved by exploiting the large data sample of \pp collisions at \thirteensix being collected during the Run~3 of the LHC by the ALICE experiment, which underwent a significant upgrade during the LHC Long Shutdown 2.


\newenvironment{acknowledgement}{\relax}{\relax}
\begin{acknowledgement}
\section*{Acknowledgements}


The ALICE Collaboration would like to thank all its engineers and technicians for their invaluable contributions to the construction of the experiment and the CERN accelerator teams for the outstanding performance of the LHC complex.
The ALICE Collaboration gratefully acknowledges the resources and support provided by all Grid centres and the Worldwide LHC Computing Grid (WLCG) collaboration.
The ALICE Collaboration acknowledges the following funding agencies for their support in building and running the ALICE detector:
A. I. Alikhanyan National Science Laboratory (Yerevan Physics Institute) Foundation (ANSL), State Committee of Science and World Federation of Scientists (WFS), Armenia;
Austrian Academy of Sciences, Austrian Science Fund (FWF): [M 2467-N36] and Nationalstiftung f\"{u}r Forschung, Technologie und Entwicklung, Austria;
Ministry of Communications and High Technologies, National Nuclear Research Center, Azerbaijan;
Rede Nacional de Física de Altas Energias (Renafae), Financiadora de Estudos e Projetos (Finep), Funda\c{c}\~{a}o de Amparo \`{a} Pesquisa do Estado de S\~{a}o Paulo (FAPESP) and The Sao Paulo Research Foundation  (FAPESP), Brazil;
Bulgarian Ministry of Education and Science, within the National Roadmap for Research Infrastructures 2020-2027 (object CERN), Bulgaria;
Ministry of Education of China (MOEC) , Ministry of Science \& Technology of China (MSTC) and National Natural Science Foundation of China (NSFC), China;
Ministry of Science and Education and Croatian Science Foundation, Croatia;
Centro de Aplicaciones Tecnol\'{o}gicas y Desarrollo Nuclear (CEADEN), Cubaenerg\'{\i}a, Cuba;
Ministry of Education, Youth and Sports of the Czech Republic, Czech Republic;
The Danish Council for Independent Research | Natural Sciences, the VILLUM FONDEN and Danish National Research Foundation (DNRF), Denmark;
Helsinki Institute of Physics (HIP), Finland;
Commissariat \`{a} l'Energie Atomique (CEA) and Institut National de Physique Nucl\'{e}aire et de Physique des Particules (IN2P3) and Centre National de la Recherche Scientifique (CNRS), France;
Bundesministerium f\"{u}r Bildung und Forschung (BMBF) and GSI Helmholtzzentrum f\"{u}r Schwerionenforschung GmbH, Germany;
General Secretariat for Research and Technology, Ministry of Education, Research and Religions, Greece;
National Research, Development and Innovation Office, Hungary;
Department of Atomic Energy Government of India (DAE), Department of Science and Technology, Government of India (DST), University Grants Commission, Government of India (UGC) and Council of Scientific and Industrial Research (CSIR), India;
National Research and Innovation Agency - BRIN, Indonesia;
Istituto Nazionale di Fisica Nucleare (INFN), Italy;
Japanese Ministry of Education, Culture, Sports, Science and Technology (MEXT) and Japan Society for the Promotion of Science (JSPS) KAKENHI, Japan;
Consejo Nacional de Ciencia (CONACYT) y Tecnolog\'{i}a, through Fondo de Cooperaci\'{o}n Internacional en Ciencia y Tecnolog\'{i}a (FONCICYT) and Direcci\'{o}n General de Asuntos del Personal Academico (DGAPA), Mexico;
Nederlandse Organisatie voor Wetenschappelijk Onderzoek (NWO), Netherlands;
The Research Council of Norway, Norway;
Pontificia Universidad Cat\'{o}lica del Per\'{u}, Peru;
Ministry of Science and Higher Education, National Science Centre and WUT ID-UB, Poland;
Korea Institute of Science and Technology Information and National Research Foundation of Korea (NRF), Republic of Korea;
Ministry of Education and Scientific Research, Institute of Atomic Physics, Ministry of Research and Innovation and Institute of Atomic Physics and Universitatea Nationala de Stiinta si Tehnologie Politehnica Bucuresti, Romania;
Ministerstvo skolstva, vyskumu, vyvoja a mladeze SR, Slovakia;
National Research Foundation of South Africa, South Africa;
Swedish Research Council (VR) and Knut \& Alice Wallenberg Foundation (KAW), Sweden;
European Organization for Nuclear Research, Switzerland;
Suranaree University of Technology (SUT), National Science and Technology Development Agency (NSTDA) and National Science, Research and Innovation Fund (NSRF via PMU-B B05F650021), Thailand;
Turkish Energy, Nuclear and Mineral Research Agency (TENMAK), Turkey;
National Academy of  Sciences of Ukraine, Ukraine;
Science and Technology Facilities Council (STFC), United Kingdom;
National Science Foundation of the United States of America (NSF) and United States Department of Energy, Office of Nuclear Physics (DOE NP), United States of America.
In addition, individual groups or members have received support from:
Czech Science Foundation (grant no. 23-07499S), Czech Republic;
FORTE project, reg.\ no.\ CZ.02.01.01/00/22\_008/0004632, Czech Republic, co-funded by the European Union, Czech Republic;
European Research Council (grant no. 950692), European Union;
Deutsche Forschungs Gemeinschaft (DFG, German Research Foundation) ``Neutrinos and Dark Matter in Astro- and Particle Physics'' (grant no. SFB 1258), Germany;
ICSC - National Research Center for High Performance Computing, Big Data and Quantum Computing and FAIR - Future Artificial Intelligence Research, funded by the NextGenerationEU program (Italy).

\end{acknowledgement}

\bibliographystyle{utphys}   
\bibliography{bibliography}

\newpage
\appendix

\appendix


\section{Systematic uncertainties for other multiplicity classes}\label{app:systerr}

\renewcommand{\arraystretch}{1.21}
\begin{table}[h!]
    \caption{Systematic uncertainties on the corrected prompt yield measured in high-multiplicity class}
    \centering
    \begin{tabular}{c|rrr|rrrr|rrrr}
    \hline\hline
    Decay channel & \multicolumn{3}{c|}{\XicPlusToXiPiPi} & \multicolumn{4}{c|}{\XicZeroToPiXi} &
    \multicolumn{4}{c}{\XicZeroToXiEleNu} \\
    \pt (\GeVc) & 4--6 & 6--8 & 8--12    & 2--4 & 4--6 & 6--8 & 8--12    & 2--4 & 4--6 & 6--8 & 8--12 \\\hline
    Raw yield       & 5\% & 6\% & 6\%    & 10\% & 10\% & 10\% & 9\%    & 21\% & 11\% & 10\% & 5\% \\
    MC \pt shape    & 1\% & negl. & negl.    & 3\% & 1\% & negl. & negl.       & 5\% & 2\% & 2\% & 3\% \\
    MC multiplicity & negl. & negl. & negl.    & 4\% & 1\% & negl. & negl.       & 5\% & 4\% & 2\% & 5\% \\
    Prompt fraction & $^{+4}_{-6}$\% & $^{+4}_{-6}$\% & $^{+6}_{-8}$\%
                    & $^{+3}_{-3}$\% & $^{+3}_{-3}$\% & $^{+4}_{-4}$\% & $^{+4}_{-4}$\%
                    & $^{+1}_{\;\;\;\,0}$\% & $^{+1}_{\;\;\;\,0}$\% & $^{+4}_{\;\;\;\,0}$\% & $^{+4}_{\;\;\;\,0}$\% \\
    Efficiency correction  & 12\% & 11\% & 6\%    & 6\% & 6\% & 6\% & 6\%   & 5\% & 5\% & 5\% & 5\% \\
    Tracking        & 5\% & 5\% & 5\%       & 5\% & 5\% & 5\% & 5\%    & 5\% & 5\% & 5\% & 5\% \\
    Unfolding       & -- & -- & --             & -- & -- & -- & --            & 2\% & 2\% & 2\% & 1\% \\
    $\left | y \right |$ variation   & -- & -- & --    & -- & -- & -- & --    & 1\% & 1\% & 1\% & 1\% 
    \\\hline
    Total & 16\% & 15\% & 12\%    & 14\% & 13\% & 13\% & 13\%   & 23\% & 14\% & 13\% & 12\% \\    
    \hline\hline
    \end{tabular}
    \label{tab:systErrHM}

\vspace{1.3\columnsep}

    \centering
    \caption{Systematic uncertainties on the corrected prompt yield measured in intermediate-multiplicity class}
    \begin{tabular}{c|rrr|rrrr|rrrr}
    \hline\hline
    Decay channel & \multicolumn{3}{c|}{\XicPlusToXiPiPi} & \multicolumn{4}{c|}{\XicZeroToPiXi} &
    \multicolumn{4}{c}{\XicZeroToXiEleNu} \\
    \pt (\GeVc) & 4--6 & 6--8 & 8--12    & 2--4 & 4--6 & 6--8 & 8--12    & 2--4 & 4--6 & 6--8 & 8--12 \\\hline
    Raw yield       & 6\% & 6\% & 6\%    & 7\% & 8\% & 9\% & 10\%  & 21\% & 11\% & 10\% & 5\% \\
    MC \pt shape    & 1\% & negl. & negl.    & 3\% & 1\% & negl. & negl.    & 5\% & 2\% & 1\% & 2\% \\
    MC multiplicity & negl. & negl. & negl.    & negl. & negl. & negl. & negl.    & 1\% & 1\% & 1\% & 2\% \\
    Prompt fraction & $^{+4}_{-6}$\% & $^{+4}_{-6}$\% & $^{+6}_{-7}$\%
                    & $^{+3}_{-2}$\% & $^{+3}_{-2}$\% & $^{+4}_{-3}$\% & $^{+4}_{-3}$\%
                    & $^{+5}_{-1}$\% & $^{+6}_{-1}$\% & $^{+4}_{-1}$\% & $^{+4}_{-3}$\% \\
    Efficiency correction  & 11\% & 11\% & 12\%    & 6\% & 6\% & 6\% & 6\%    & 7\% & 6\% & 6\% & 6\% \\
    Tracking        & 5\% & 5\% & 5\%       & 5\% & 5\% & 5\% & 5\%    & 5\% & 5\% & 5\% & 5\% \\
    Unfolding       & -- & -- & --             & -- & -- & -- & --            & 1\% & 4\% & 3\% & 4\% \\
    $\left | y \right |$ variation   & -- & -- & --    & -- & -- & -- & --    & 1\% & 1\% & 1\% & 1\%
    \\\hline
    Total & 14\% & 15\% & 16\%    & 11\% & 11\% & 12\% & 13\%   & 23\% & 15\% & 14\% & 12\% \\    
    \hline\hline
    \end{tabular}
    \label{tab:systErrMB0p1to30}

\vspace{1.3\columnsep}

    \centering
    \caption{Systematic uncertainties on the corrected prompt yield measured in low-multiplicity class}
    \begin{tabular}{c|rr|rrr|rrr}
    \hline\hline
    Decay channel & \multicolumn{2}{c|}{\XicPlusToXiPiPi} & \multicolumn{3}{c|}{\XicZeroToPiXi} &
    \multicolumn{3}{c}{\XicZeroToXiEleNu} \\
    \pt (\GeVc) & 4--6 & 6--8    & 2--4 & 4--6 & 6--8    & 2--4 & 4--6 & 6--8 \\\hline
    Raw yield & 5\% & 8\%     & 10\% & 10\% & 9\%        & 21\% & 11\% & 10\% \\
    MC \pt shape    & 1\% & negl.     & 3\% & 1\% & negl.         & 5\% & 2\% & 1\% \\
    MC multiplicity & negl. & negl.     & negl. & negl. & negl.         & 1\% & 1\% & 1\% \\
    Prompt fraction & $^{+4}_{-6}$\% & $^{+4}_{-6}$\%
                    & $^{+3}_{-2}$\% & $^{+3}_{-2}$\% & $^{+4}_{-3}$\%
                    & $^{+5}_{-1}$\% & $^{+6}_{-1}$\% & $^{+4}_{-1}$\% \\
    Efficiency correction  & 6\% & 14\%    & 10\% & 10\% & 10\%       & 7\% & 6\% & 6\% \\
    Tracking        & 5\% & 5\%     & 5\% & 5\% & 5\%         & 5\% & 5\% & 5\% \\   
    Unfolding       & -- & --         & -- & -- & --               & 1\% & 4\% & 3\% \\    
    $\left | y \right |$ variation  & -- & --     & -- & -- & --   & 1\% & 1\% & 1\% 
    \\\hline
    Total & 11\% & 18\%     & 16\% & 15\% & 15\%    & 23\% & 15\% & 14\% \\
    \hline\hline
    \end{tabular}
    \label{tab:systErrMB30to100}
\end{table}
\clearpage 

\section{The ALICE Collaboration}
\label{app:collab}
\begin{flushleft} 
\small

I.J.~Abualrob\,\orcidlink{0009-0005-3519-5631}\,$^{\rm 114}$, 
S.~Acharya\,\orcidlink{0000-0002-9213-5329}\,$^{\rm 50}$, 
G.~Aglieri Rinella\,\orcidlink{0000-0002-9611-3696}\,$^{\rm 32}$, 
L.~Aglietta\,\orcidlink{0009-0003-0763-6802}\,$^{\rm 24}$, 
M.~Agnello\,\orcidlink{0000-0002-0760-5075}\,$^{\rm 29}$, 
N.~Agrawal\,\orcidlink{0000-0003-0348-9836}\,$^{\rm 25}$, 
Z.~Ahammed\,\orcidlink{0000-0001-5241-7412}\,$^{\rm 133}$, 
S.~Ahmad\,\orcidlink{0000-0003-0497-5705}\,$^{\rm 15}$, 
I.~Ahuja\,\orcidlink{0000-0002-4417-1392}\,$^{\rm 36}$, 
ZUL.~Akbar$^{\rm 81}$, 
A.~Akindinov\,\orcidlink{0000-0002-7388-3022}\,$^{\rm 139}$
% ...

\section*{Affiliation Notes}

$^{\rm I}$ Deceased\\
$^{\rm II}$ Also at: Max-Planck-Institut fur Physik, Munich, Germany\\
$^{\rm III}$ Also at: Czech Technical University in Prague (CZ)\\
$^{\rm IV}$ Also at: Italian National Agency for New Technologies, Energy and Sustainable Economic Development (ENEA), Bologna, Italy\\
$^{\rm V}$ Also at: Instituto de Fisica da Universidade de Sao Paulo\\
$^{\rm VI}$ Also at: Dipartimento DET del Politecnico di Torino, Turin, Italy\\
$^{\rm VII}$ Also at: Department of Applied Physics, Aligarh Muslim University, Aligarh, India\\
$^{\rm VIII}$ Also at: Institute of Theoretical Physics, University of Wroclaw, Poland\\
$^{\rm IX}$ Also at: Facultad de Ciencias, Universidad Nacional Aut\'{o}noma de M\'{e}xico, Mexico City, Mexico\\

\section*{Collaboration Institutes}

$^{1}$ A.I. Alikhanyan National Science Laboratory (Yerevan Physics Institute) Foundation, Yerevan, Armenia\\
$^{2}$ AGH University of Krakow, Cracow, Poland\\
$^{3}$ Bogolyubov Institute for Theoretical Physics, National Academy of Sciences of Ukraine, Kyiv, Ukraine\\
$^{4}$ Bose Institute, Department of Physics  and Centre for Astroparticle Physics and Space Science (CAPSS), Kolkata, India\\
$^{5}$ California Polytechnic State University, San Luis Obispo, California, United States\\
$^{6}$ Central China Normal University, Wuhan, China\\
$^{7}$ Centro de Aplicaciones Tecnol\'{o}gicas y Desarrollo Nuclear (CEADEN), Havana, Cuba\\
$^{8}$ Centro de Investigaci\'{o}n y de Estudios Avanzados (CINVESTAV), Mexico City and M\'{e}rida, Mexico\\
$^{9}$ Chicago State University, Chicago, Illinois, United States\\
$^{10}$ China Nuclear Data Center, China Institute of Atomic Energy, Beijing, China\\
$^{11}$ China University of Geosciences, Wuhan, China\\
$^{12}$ Chungbuk National University, Cheongju, Republic of Korea\\
$^{13}$ Comenius University Bratislava, Faculty of Mathematics, Physics and Informatics, Bratislava, Slovak Republic\\
$^{14}$ Creighton University, Omaha, Nebraska, United States\\
$^{15}$ Department of Physics, Aligarh Muslim University, Aligarh, India\\
$^{16}$ Department of Physics, Pusan National University, Pusan, Republic of Korea\\
$^{17}$ Department of Physics, Sejong University, Seoul, Republic of Korea\\
$^{18}$ Department of Physics, University of California, Berkeley, California, United States\\
$^{19}$ Department of Physics, University of Oslo, Oslo, Norway\\
$^{20}$ Department of Physics and Technology, University of Bergen, Bergen, Norway\\
$^{21}$ Dipartimento di Fisica, Universit\`{a} di Pavia, Pavia, Italy\\
$^{22}$ Dipartimento di Fisica dell'Universit\`{a} and Sezione INFN, Cagliari, Italy\\
$^{23}$ Dipartimento di Fisica dell'Universit\`{a} and Sezione INFN, Trieste, Italy\\
$^{24}$ Dipartimento di Fisica dell'Universit\`{a} and Sezione INFN, Turin, Italy\\
$^{25}$ Dipartimento di Fisica e Astronomia dell'Universit\`{a} and Sezione INFN, Bologna, Italy\\
$^{26}$ Dipartimento di Fisica e Astronomia dell'Universit\`{a} and Sezione INFN, Catania, Italy\\
$^{27}$ Dipartimento di Fisica e Astronomia dell'Universit\`{a} and Sezione INFN, Padova, Italy\\
$^{28}$ Dipartimento di Fisica `E.R.~Caianiello' dell'Universit\`{a} and Gruppo Collegato INFN, Salerno, Italy\\
$^{29}$ Dipartimento DISAT del Politecnico and Sezione INFN, Turin, Italy\\
$^{30}$ Dipartimento di Scienze MIFT, Universit\`{a} di Messina, Messina, Italy\\
$^{31}$ Dipartimento Interateneo di Fisica `M.~Merlin' and Sezione INFN, Bari, Italy\\
$^{32}$ European Organization for Nuclear Research (CERN), Geneva, Switzerland\\
$^{33}$ Faculty of Electrical Engineering, Mechanical Engineering and Naval Architecture, University of Split, Split, Croatia\\
$^{34}$ Faculty of Nuclear Sciences and Physical Engineering, Czech Technical University in Prague, Prague, Czech Republic\\
$^{35}$ Faculty of Physics, Sofia University, Sofia, Bulgaria\\
$^{36}$ Faculty of Science, P.J.~\v{S}af\'{a}rik University, Ko\v{s}ice, Slovak Republic\\
$^{37}$ Faculty of Technology, Environmental and Social Sciences, Bergen, Norway\\
$^{38}$ Frankfurt Institute for Advanced Studies, Johann Wolfgang Goethe-Universit\"{a}t Frankfurt, Frankfurt, Germany\\
$^{39}$ Fudan University, Shanghai, China\\
$^{40}$ Gangneung-Wonju National University, Gangneung, Republic of Korea\\
$^{41}$ Gauhati University, Department of Physics, Guwahati, India\\
$^{42}$ Helmholtz-Institut f\"{u}r Strahlen- und Kernphysik, Rheinische Friedrich-Wilhelms-Universit\"{a}t Bonn, Bonn, Germany\\
$^{43}$ Helsinki Institute of Physics (HIP), Helsinki, Finland\\
$^{44}$ High Energy Physics Group,  Universidad Aut\'{o}noma de Puebla, Puebla, Mexico\\
$^{45}$ Horia Hulubei National Institute of Physics and Nuclear Engineering, Bucharest, Romania\\
$^{46}$ HUN-REN Wigner Research Centre for Physics, Budapest, Hungary\\
$^{47}$ Indian Institute of Technology Bombay (IIT), Mumbai, India\\
$^{48}$ Indian Institute of Technology Indore, Indore, India\\
$^{49}$ INFN, Laboratori Nazionali di Frascati, Frascati, Italy\\
$^{50}$ INFN, Sezione di Bari, Bari, Italy\\
$^{51}$ INFN, Sezione di Bologna, Bologna, Italy\\
$^{52}$ INFN, Sezione di Cagliari, Cagliari, Italy\\
$^{53}$ INFN, Sezione di Catania, Catania, Italy\\
$^{54}$ INFN, Sezione di Padova, Padova, Italy\\
$^{55}$ INFN, Sezione di Pavia, Pavia, Italy\\
$^{56}$ INFN, Sezione di Torino, Turin, Italy\\
$^{57}$ INFN, Sezione di Trieste, Trieste, Italy\\
$^{58}$ Inha University, Incheon, Republic of Korea\\
$^{59}$ Institute for Gravitational and Subatomic Physics (GRASP), Utrecht University/Nikhef, Utrecht, Netherlands\\
$^{60}$ Institute of Experimental Physics, Slovak Academy of Sciences, Ko\v{s}ice, Slovak Republic\\
$^{61}$ Institute of Physics, Homi Bhabha National Institute, Bhubaneswar, India\\
$^{62}$ Institute of Physics of the Czech Academy of Sciences, Prague, Czech Republic\\
$^{63}$ Institute of Space Science (ISS), Bucharest, Romania\\
$^{64}$ Institut f\"{u}r Kernphysik, Johann Wolfgang Goethe-Universit\"{a}t Frankfurt, Frankfurt, Germany\\
$^{65}$ Instituto de Ciencias Nucleares, Universidad Nacional Aut\'{o}noma de M\'{e}xico, Mexico City, Mexico\\
$^{66}$ Instituto de F\'{i}sica, Universidade Federal do Rio Grande do Sul (UFRGS), Porto Alegre, Brazil\\
$^{67}$ Instituto de F\'{\i}sica, Universidad Nacional Aut\'{o}noma de M\'{e}xico, Mexico City, Mexico\\
$^{68}$ iThemba LABS, National Research Foundation, Somerset West, South Africa\\
$^{69}$ Jeonbuk National University, Jeonju, Republic of Korea\\
$^{70}$ Johann-Wolfgang-Goethe Universit\"{a}t Frankfurt Institut f\"{u}r Informatik, Fachbereich Informatik und Mathematik, Frankfurt, Germany\\
$^{71}$ Korea Institute of Science and Technology Information, Daejeon, Republic of Korea\\
$^{72}$ Laboratoire de Physique Subatomique et de Cosmologie, Universit\'{e} Grenoble-Alpes, CNRS-IN2P3, Grenoble, France\\
$^{73}$ Lawrence Berkeley National Laboratory, Berkeley, California, United States\\
$^{74}$ Lund University Department of Physics, Division of Particle Physics, Lund, Sweden\\
$^{75}$ Nagasaki Institute of Applied Science, Nagasaki, Japan\\
$^{76}$ Nara Women{'}s University (NWU), Nara, Japan\\
$^{77}$ National and Kapodistrian University of Athens, School of Science, Department of Physics , Athens, Greece\\
$^{78}$ National Centre for Nuclear Research, Warsaw, Poland\\
$^{79}$ National Institute of Science Education and Research, Homi Bhabha National Institute, Jatni, India\\
$^{80}$ National Nuclear Research Center, Baku, Azerbaijan\\
$^{81}$ National Research and Innovation Agency - BRIN, Jakarta, Indonesia\\
$^{82}$ Niels Bohr Institute, University of Copenhagen, Copenhagen, Denmark\\
$^{83}$ Nikhef, National institute for subatomic physics, Amsterdam, Netherlands\\
$^{84}$ Nuclear Physics Group, STFC Daresbury Laboratory, Daresbury, United Kingdom\\
$^{85}$ Nuclear Physics Institute of the Czech Academy of Sciences, Husinec-\v{R}e\v{z}, Czech Republic\\
$^{86}$ Oak Ridge National Laboratory, Oak Ridge, Tennessee, United States\\
$^{87}$ Ohio State University, Columbus, Ohio, United States\\
$^{88}$ Physics department, Faculty of science, University of Zagreb, Zagreb, Croatia\\
$^{89}$ Physics Department, Panjab University, Chandigarh, India\\
$^{90}$ Physics Department, University of Jammu, Jammu, India\\
$^{91}$ Physics Program and International Institute for Sustainability with Knotted Chiral Meta Matter (WPI-SKCM$^{2}$), Hiroshima University, Hiroshima, Japan\\
$^{92}$ Physikalisches Institut, Eberhard-Karls-Universit\"{a}t T\"{u}bingen, T\"{u}bingen, Germany\\
$^{93}$ Physikalisches Institut, Ruprecht-Karls-Universit\"{a}t Heidelberg, Heidelberg, Germany\\
$^{94}$ Physik Department, Technische Universit\"{a}t M\"{u}nchen, Munich, Germany\\
$^{95}$ Politecnico di Bari and Sezione INFN, Bari, Italy\\
$^{96}$ Research Division and ExtreMe Matter Institute EMMI, GSI Helmholtzzentrum f\"ur Schwerionenforschung GmbH, Darmstadt, Germany\\
$^{97}$ Saga University, Saga, Japan\\
$^{98}$ Saha Institute of Nuclear Physics, Homi Bhabha National Institute, Kolkata, India\\
$^{99}$ School of Physics and Astronomy, University of Birmingham, Birmingham, United Kingdom\\
$^{100}$ Secci\'{o}n F\'{\i}sica, Departamento de Ciencias, Pontificia Universidad Cat\'{o}lica del Per\'{u}, Lima, Peru\\
$^{101}$ Stefan Meyer Institut f\"{u}r Subatomare Physik (SMI), Vienna, Austria\\
$^{102}$ SUBATECH, IMT Atlantique, Nantes Universit\'{e}, CNRS-IN2P3, Nantes, France\\
$^{103}$ Sungkyunkwan University, Suwon City, Republic of Korea\\
$^{104}$ Suranaree University of Technology, Nakhon Ratchasima, Thailand\\
$^{105}$ Technical University of Ko\v{s}ice, Ko\v{s}ice, Slovak Republic\\
$^{106}$ The Henryk Niewodniczanski Institute of Nuclear Physics, Polish Academy of Sciences, Cracow, Poland\\
$^{107}$ The University of Texas at Austin, Austin, Texas, United States\\
$^{108}$ Universidad Aut\'{o}noma de Sinaloa, Culiac\'{a}n, Mexico\\
$^{109}$ Universidade de S\~{a}o Paulo (USP), S\~{a}o Paulo, Brazil\\
$^{110}$ Universidade Estadual de Campinas (UNICAMP), Campinas, Brazil\\
$^{111}$ Universidade Federal do ABC, Santo Andre, Brazil\\
$^{112}$ Universitatea Nationala de Stiinta si Tehnologie Politehnica Bucuresti, Bucharest, Romania\\
$^{113}$ University of Derby, Derby, United Kingdom\\
$^{114}$ University of Houston, Houston, Texas, United States\\
$^{115}$ University of Jyv\"{a}skyl\"{a}, Jyv\"{a}skyl\"{a}, Finland\\
$^{116}$ University of Kansas, Lawrence, Kansas, United States\\
$^{117}$ University of Liverpool, Liverpool, United Kingdom\\
$^{118}$ University of Science and Technology of China, Hefei, China\\
$^{119}$ University of South-Eastern Norway, Kongsberg, Norway\\
$^{120}$ University of Tennessee, Knoxville, Tennessee, United States\\
$^{121}$ University of the Witwatersrand, Johannesburg, South Africa\\
$^{122}$ University of Tokyo, Tokyo, Japan\\
$^{123}$ University of Tsukuba, Tsukuba, Japan\\
$^{124}$ Universit\"{a}t M\"{u}nster, Institut f\"{u}r Kernphysik, M\"{u}nster, Germany\\
$^{125}$ Universit\'{e} Clermont Auvergne, CNRS/IN2P3, LPC, Clermont-Ferrand, France\\
$^{126}$ Universit\'{e} de Lyon, CNRS/IN2P3, Institut de Physique des 2 Infinis de Lyon, Lyon, France\\
$^{127}$ Universit\'{e} de Strasbourg, CNRS, IPHC UMR 7178, F-67000 Strasbourg, France, Strasbourg, France\\
$^{128}$ Universit\'{e} Paris-Saclay, Centre d'Etudes de Saclay (CEA), IRFU, D\'{e}partment de Physique Nucl\'{e}aire (DPhN), Saclay, France\\
$^{129}$ Universit\'{e}  Paris-Saclay, CNRS/IN2P3, IJCLab, Orsay, France\\
$^{130}$ Universit\`{a} degli Studi di Foggia, Foggia, Italy\\
$^{131}$ Universit\`{a} del Piemonte Orientale, Vercelli, Italy\\
$^{132}$ Universit\`{a} di Brescia, Brescia, Italy\\
$^{133}$ Variable Energy Cyclotron Centre, Homi Bhabha National Institute, Kolkata, India\\
$^{134}$ Warsaw University of Technology, Warsaw, Poland\\
$^{135}$ Wayne State University, Detroit, Michigan, United States\\
$^{136}$ Yale University, New Haven, Connecticut, United States\\
$^{137}$ Yildiz Technical University, Istanbul, Turkey\\
$^{138}$ Yonsei University, Seoul, Republic of Korea\\
$^{139}$ Affiliated with an institute formerly covered by a cooperation agreement with CERN\\
$^{140}$ Affiliated with an international laboratory covered by a cooperation agreement with CERN.\\

\end{flushleft} 
 
\end{document}